Title: Beam dynamics for the Scorpius Conceptual Design Report

Author(s): Ekdahl, Carl August Jr.

Intended for: Report

Issued: 2017-10-06 (Draft)

# Beam Dynamics for the Scorpius Conceptual Design Report

Carl Ekdahl

## I. Introduction

Scorpius is a new multi-pulse linear induction accelerator (LIA). It is presently planned for Scorpius to accelerate four 2-kA electron pulses to 20 MeV from an injected energy of 2 MeV. The goal of Scorpius is to deliver a multi-pulse electron beam with quality sufficient for use in flash radiography of large-scale explosively-driven experiments. Beam physics has a profound effect on the quality of multi-pulse radiographs. Beam halo or asymmetry causes emittance growth, which enlarges spot size, thereby degrading resolution. Beam centroid motion due to instabilities is especially problematic. High-frequency motion blurs the radiographic spot size, thereby degrading resolution. Low-frequency motion causes pulse-to-pulse wandering of the source spot, thereby degrading registration of successive radiographs. We use a wide range of tools to investigate potential beam physics problems associated with Scorpius. In addition to analytic theory and experimental data from our operational LIAs, we use a number of reliable beam dynamics computer codes. These include both beam envelope and particle in cell (PIC) codes. These codes are described in section VII.

The most dangerous instability for electron linacs is the beam breakup (BBU) instability [1, 2, 3]. For radiography LIAs it is particularly troublesome, because even if it is not strong enough to destroy the beam, the high-frequency BBU motion can blur the source spot , which is time-integrated over the pulselength. In operational practice, the BBU is suppressed by the magnetic focusing fields used to transport the beam through the LIA. Therefore, the beam dynamics strategy that we use for Scorpius is to first design a magnetic field-strength profile that will adequately suppress the BBU, and then demonstrate that this profile (the so-called "tune") does not lead to other problems. The following sections describe each potential beam physics problem, and its resolution.

## II. Matched Transport

The electron beam is transported through the Scorpius LIA using solenoidal magnetic focusing fields. This is an efficient and convenient means that has been used in all electron LIAs since the very first. Each accelerating cell has a solenoid incorporated into it, as well as dipole windings for steering. The magnetic field produced by these magnets is called the "tune" of the accelerator. This section reports the results of beam simulations of tunes for Scorpius. These calculations use external magnetic and electric fields resulting from simulations of the Scorpius induction-cell conceptual design (see Section VII.A). There are 72 accelerator cells required to reach the 20-MeV final energy if a 2-MeV injector is used.

We use the XTR envelope code [4]to design tunes for DARHT. The design goal is a tune that adequately suppresses BBU. The exponential growth factor for BBU is proportional to $\langle 1/B \rangle$, where $B$ is the magnetic field on axis provided by the focusing solenoids used to transport the beam through the accelerator. Thus, a magnetic field tune designed to reduce spot size blur from BBU should minimize $\langle 1/B \rangle$, taking into account any other constraints. For example, Caporaso showed that a tune with magnetic field increasing as $\sqrt{\gamma}$ would minimize the corkscrew motion for any given BBU amplification [5]. However, since corkscrew motion can be effectively reduced with corrector dipoles [6], other constraints have become more important. Among these are magnet power requirements and magnet heating. Thus, we have considered a class of magnet-constrained tunes for Scorpius. These tunes are constrained by an initial field $B_0$ at the injector end that is high enough to suppress the image displacement instability, and a final field $B_f$ at the LIA exit that is limited by solenoid heating. For example, consider a tune with continuous magnetic field increasing as

$$B(\gamma) = B_0 + (B_f - B_0) \left( \frac{\gamma - \gamma_0}{\gamma_f - \gamma_0} \right)^p \tag{1}$$

For a constant accelerating gradient; $\gamma = \gamma_0 + z \, d\gamma / dz$ with $\Delta B = B_f - B_0$ this reduces to

$$B(z) = B_0 + \Delta B \left( z / L \right)^p \tag{2}$$

Figure 1 shows this tune profile for three different values of $p$ and Figure 2 shows how much $\langle 1/B \rangle$ can be reduced to suppress BBU for $B_0 = 200 \, \text{G}$ and $B_f = 1.5 \, \text{kG}$. With these magnet constraints it is clear that $\langle 1/B \rangle$ is limited to values greater than unity. The tuning strategy that we used on DARHT-II was to begin with $p$=0.5 to minimize corkscrew, and then adjust downward to further suppress BBU as needed. We used the same strategy for the Scorpius tune for this report.

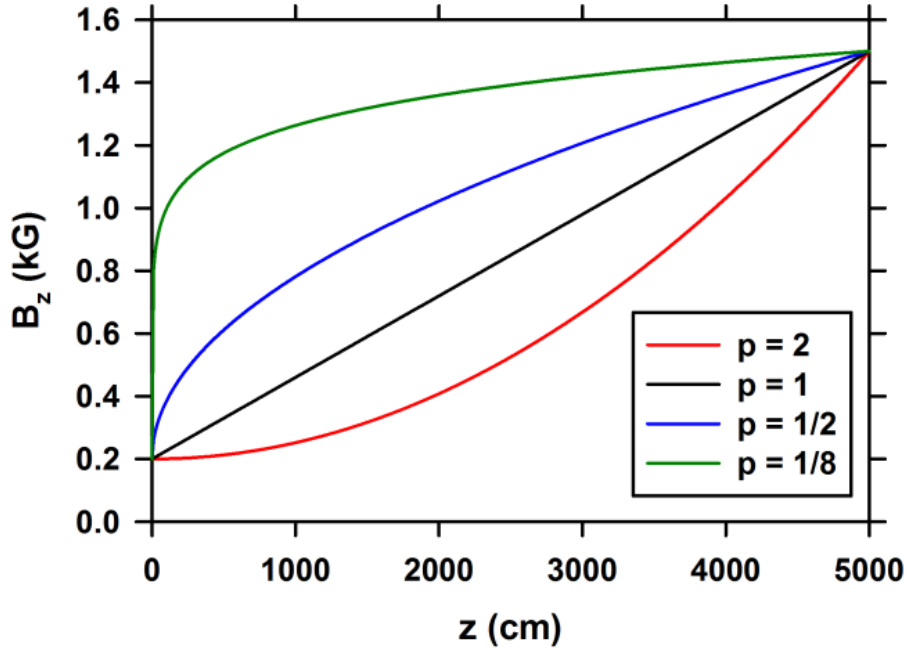

Figure 1 Continuous tune profiles for varying exponents.

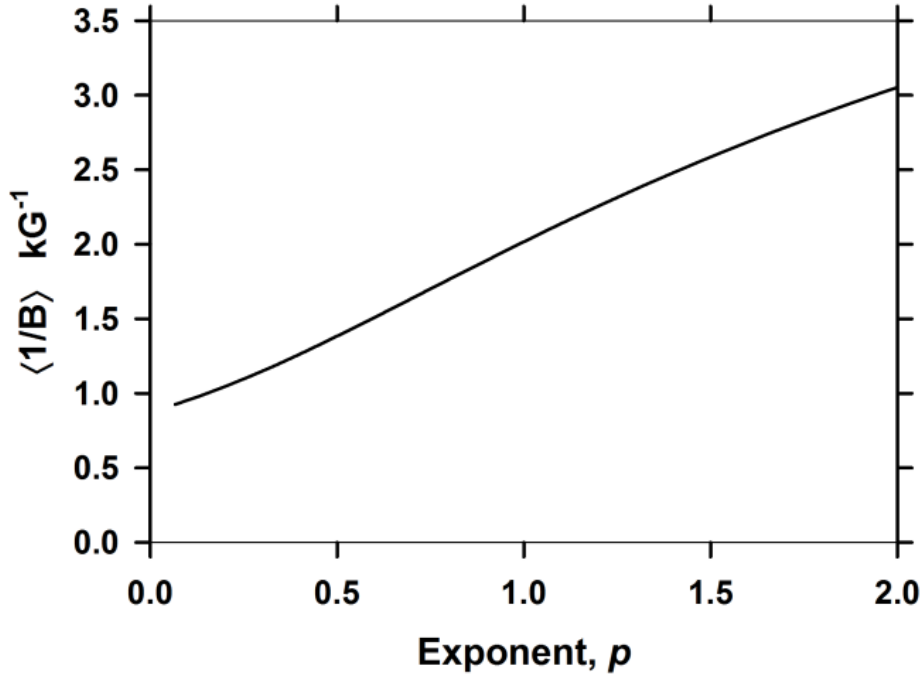

**Figure 2: Average $\langle 1/B \rangle$ for the tune described by Eq. (2) showing the advantage of reducing p in order to reduce $\langle 1/B \rangle$, thereby further suppressing BBU.**

The magnetic fields produced by the Scorpius magnets are limited to less than 2-kG in order to limit the temperature rise to less than 20C. Except for the "pitch-and-catch" solenoids used to transport across gaps between cell blocks, we constrained the peak fields of the Scorpius tune to be less than 1.5 kG for this Conceptual Design Report (CDR). Moreover, we constrained the peak fields in the first cell block to be greater than ~ 200 G in order to prevent the image displacement instability (IDI). We initially started with a peak field following the p= ½ profile as described above, and then adjusted the first few magnets to match the injected beam to this profile for the remainder of the LIA. The injected beam parameters for this report were provided by the designers of the injector [7], and are shown in Table 1. The CDR tune and the resulting beam envelope calculated by the XTR envelope code (see section VII) are shown in Figure 3.

**Table 1. Initial parameters from AMBER injector calculations [7].**

| Parameter | Symbol | Units | Value |
|---|---|---|---|
| Beam Energy | $E_0$ | MeV | 2.1 |
| Beam Current | $I_b$ | kA | 2.07 |
| Envelope Radius | $r_0$ | cm | 4.43 |
| Envelope Convergence | $r'_0$ | mr | 26.31 |
| Normalized emittance | $\varepsilon_n$ | mm-mr | 305 |

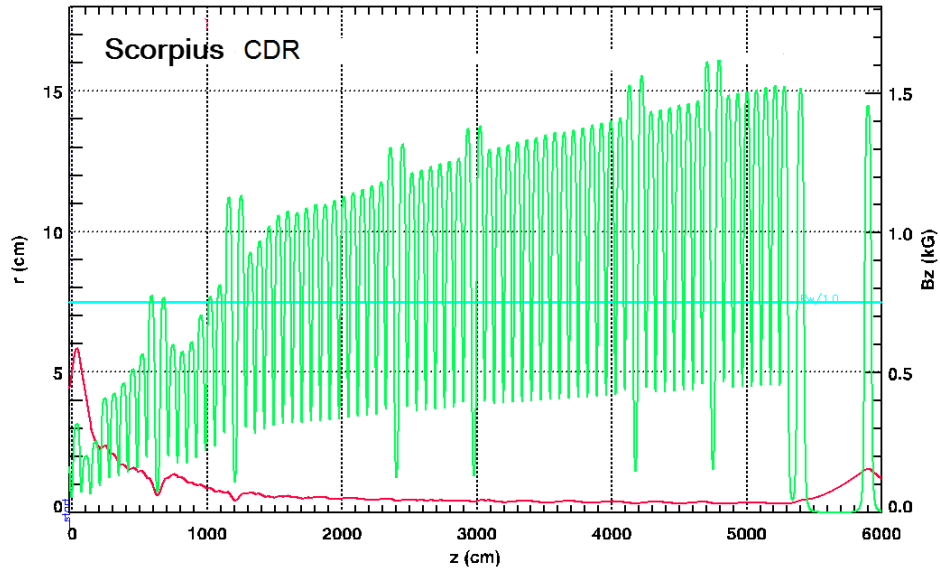

**Figure 3: Scorpius CDR tune magnetic field (green) and beam envelope radius calculated by XTR (red).**

## III. Emittance Growth

We calculated the beam emittance for the Scorpius CDR tune transport using the LSP-Slice PIC code (see Section VII).There is no inherent emittance growth in this tune if the beam initial parameters are matched as shown in Figure 3. This will require careful measurement of the injected beam parameters, followed by a detailed retuning of the first few cells of the LIA. Even for a slightly mismatched beam, with slight envelope oscillations, there is no emittance growth (see Figure 4). This thresholded onset is characteristic of emittance growth caused by halo formation [8].

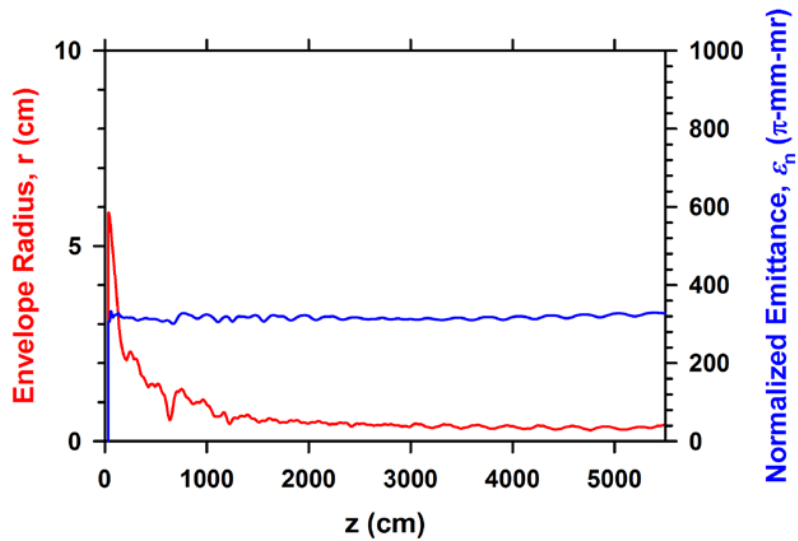

**Figure 4: PIC code simulation of beam emittance for a beam that is almost matched to the tune. Red curve: matched envelope radius. Blue curve: Beam emittance showing no growth.**

## A. Spherical aberration

A well-known contributor to emittance growth in solenoidal focusing systems is spherical aberration [9, 10, 11], which over-focuses the edge of the beam, producing hollow beam profiles. The magnetic fields used for these simulations include these aberrations, and edge-focusing is noticeable in the PIC simulations. However, for a well-matched beam, as illustrated in Figure 4, the cumulative effect of edge focusing is negligible.

Since emittance growth due to spherical aberration of a solenoid lens is proportional to the fourth power of the beam size [11], we design the tunes for DARHT-II to rapidly focus the beam to a small size. Even though the cumulative spherical aberration is noticeable in the baseline simulations (Fig.4) there is apparently little emittance growth (< ~10 π-mm-mr) due to this effect in our baseline simulation (Fig. 3). The effect is mostly due to launching the PIC slice at the diode exit, where the beam is large.

## B. Mismatched beam

Emittance growth can result from envelope oscillations caused by a mismatch of the beam to the magnetic transport system. A badly mismatched beam exhibits large envelope oscillations, sometimes called a "sausage," "m=0," or "breathing" mode. The mechanism of this contribution to emittance growth is parametric amplification of electron orbits that resonate with the envelope oscillation, expelling those electrons from the beam core into a halo [12, 13].

Figure 5 shows the envelope oscillations of a mismatched beam and the resulting emittance growth in an LSP-slice simulation [14]. Halo growth was quite clear in LSP-slice movies of the beam distribution as it propagated through the LIA. Figure 5 shows how the emittance grows if the actual beam injection energy were 5% less than the value used to design the tune, like could be caused by poorly calibrated diagnostics. It is noteworthy that the emittance growth saturates due to damping of the envelope oscillations.

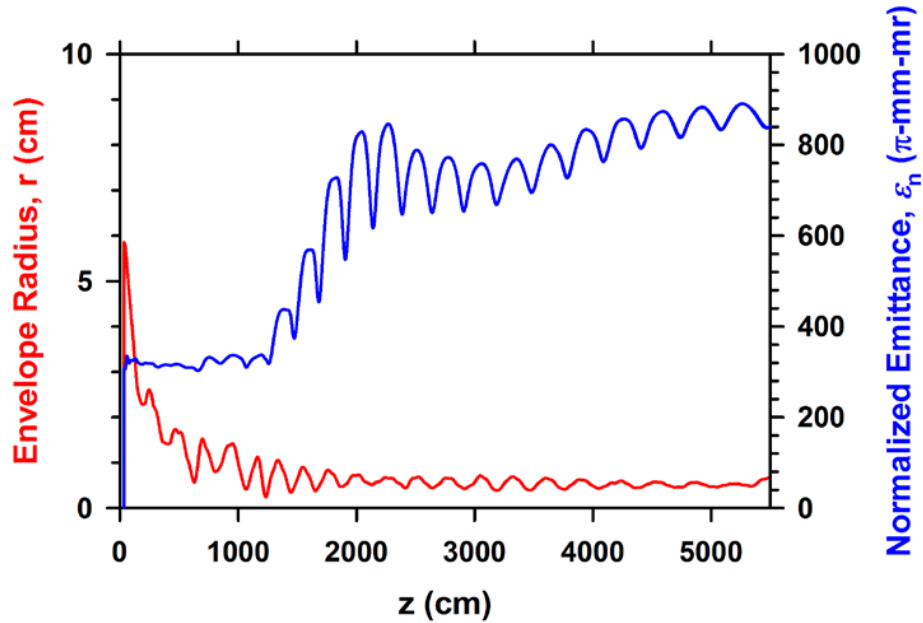

Figure 5: PIC code simulation of beam emittance for a beam that is not well-matched to the tune. Red curve: mismatched envelope radius with large envelope oscillations. Blue curve: Beam emittance showing significant growth.

Detailed PIC code simulations show that this growth is the result of beam halo generation caused by the envelope oscillations [8]. The mechanism for halo growth is that electrons in the beam see a periodic potential well due to the envelope oscillations. Electron orbits in resonance with potential well period can be parametrically amplified, thereby ejecting the particles into a halo [8, 13].

Several striking features of this mechanism are evident from PIC simulation results.

- There is a threshold of oscillation amplitude for emittance growth.
- When the initial envelope oscillations are small, the emittance grows almost linearly
- When the initial envelope oscillations are large, the emittance rapidly grows and then saturates.
- The large halo generated on these severely mismatched beams appears to damp the oscillations after the emittance saturates.
- The most severe cases show evidence of multiple halos.

Finally, beam halo is especially troublesome for radiography accelerators, such as DARHT, because the wings of the radiographic source spot caused by the halo blurs the image. The effect is as if a low-resolution image due to the halo alone were superimposed on a high-resolution image due to the core. Thus, mismatched-beam generated halo is to be prevented. Since the ideal source-spot size calculated from beam dynamics is directly proportional to emittance, and the emittance is so strongly affected by the halo, the emittance is an effective metric of this radiographic resolution degradation due to halo.

# IV.    Beam Instabilities

Beam quality can also be degraded by instabilities. Since beam quality is of paramount importance for a radiography accelerator like Scorpius, we have assessed the most dangerous instabilities for high-current LIAs. These are

- Beam Breakup (BBU)
- Image Displacement (IDI)
- Diocotron
- Resistive Wall
- Ion Hose
- Parametric Envelope

Although corkscrew motion is not strictly an instability, we discuss it here because it is historically second only to BBU in its capability for causing mischief in LIAs. Resistive wall and ion hose instabilities are generally thought of as insignificant for short pulses, but we have shown that they can be problematic for multiple pulses, so mitigation of these deserves attention in the design of Scorpius.

## A. Beam Breakup

The BBU instability is the result of the beam deflection by transverse magnetic ($TM_{1n0}$) RF modes of the accelerating cavities, which impress an RF oscillation on the beam. For radiography LIAs BBU is particularly troublesome, because even if it is not strong enough to destroy the beam, the high-frequency motion can blur the source spot, since it is time-integrated over the pulse-width.

In an LIA, the cavities are separated by lengths of beam pipe in which the cavity TM modes are cut off, so the cavities can only communicate via beam oscillations. Each successive cavity reinforces the beam oscillation, which eventually grows exponentially under the right conditions. Beam focusing between cells by external solenoidal magnetic fields reduces the oscillation amplitude, but if focusing is not strong enough, the exponential amplification wins out.

For a large enough number of accelerating cells, theory predicts that the BBU growth asymptotes to

$$\xi(z) = \xi_0 \left[ \gamma_0 / \gamma(z) \right]^{1/2} \exp(\Gamma_m), \tag{3}$$

where subscript zero denotes initial conditions, and $\gamma$ is the relativistic mass factor. The maximum growth exponent in this equation is

$$\Gamma_m(z) = \frac{I_{kA} N_g Z_{\perp\Omega/m}}{3 \times 10^4} \left\langle \frac{1}{B_{kG}} \right\rangle, \tag{4}$$

where $I_{kA}$ is the beam current in kA, $N_g$ is the number of gaps, $Z_{\perp\Omega/m}$ is the transverse impedance in $\Omega$/m, $B_{kG}$ is the guide field in kG, and $\left\langle \; \right\rangle$ indicates averaging over $z$ [10]. This theoretical maximum amplitude of the BBU in high-current LIAs has been experimentally confirmed [15, 16], and used to design DARHT-II tunes that suppress BBU amplification to acceptable levels [17, 6, 14].

We have reexamined the BBU problem for Scorpius using the Linear Accelerator Model for DARHT (LAMDA) beam dynamics code [18, 19], along with contemporary parameters for the new LIA. LAMDA has been benchmarked against analytic theory, against the Lawrence Livermore National Laboratory (LLNL) Breakup code [20], and against experimental data [21]. We briefly describe the LAMDA BBU computational algorithm in section VII.

## *Transverse Impedance*

The Scorpius cell design is based on the proven DARHT-I cell, but with the ferrite cores replaced with Metglas to provide enough flux swing (volt-seconds) for four-pulse operation (see Figure 36). The DARHT-I cavity design has been retained as closely as possible so that the transverse impedance of Scorpius will be close to that of the measured DARHT-I impedance. With one exception, the accelerating gap, insulator, RF cavity shape, and RF cavity wall materials of the cell are identical to the DARHT-I cell in order to have the same RF properties. The exception is that the ferrite disk forming the RF cavity wall on the inductor side was reduced to only a fraction of the wall in Scorpius in order to prevent breakdown when it saturates because of the high-voltage pulse.

Since we do not yet have measurements of the transverse impedance of the Scorpius cell, the impedance we use is the measured DARHT MOD2 impedance for BBU calculations in this report. This is based on the assumption that the Scorpius cavity geometry is close enough to the MOD2 geometry that there will be negligible difference in impedances. The provenance of this assumption is provided by AMOS calculations; both early comparisons with measurements of DARHT cavities [22, 23], and more recent extrapolations to Scorpius geometries [24]. The most relevant results of these are summarized in Table 2. These AMOS results support the following suppositions:

- The DARHT MOD2 impedance is nearly the same as the MOD2A impedance.
- The DARHT MOD2A impedance is nearly the same as the Scorpius impedance (with either style ferrite disk)
- Therefore, the measured DARHT MOD2 impedance is a reasonable facsimile for Scorpius BBU calculations.

**Table 2. Resonance Parameters from Measurements and AMOS calculations**

| Resonance: | 1 | | 2 | | 3 | |
|---|---|---|---|---|---|---|
| | f MHz | max $Z_\perp$ $\Omega$/cm | f MHz | max $Z_\perp$ $\Omega$/cm | f MHz | max $Z_\perp$ $\Omega$/cm |
| AMOS [24] | | | | | | |

| | | | | | |
|---|---|---|---|---|---|
| DARHT MOD2 | 279 | 6.38 | 788 | 9.35 | | |
| DARHT MOD2A | 283 | 7.12 | 788 | 9.84 | 1361 | 2.26 |
| Scorpius    (Half Disk) | 283 | 8.36 | 779 | 8.73 | 1430 | 4.89 |
| Scorpius  (Segmented Disk) | 289 | 7.46 | 780 | 10.04 | 1553 | 2.77 |
| **Measured** [22] | | | | | | |
| DARHT MOD2   (no rods) | 297 | 6.37 | 822 | 10.01 | n/a | n/a |
| DARHT MOD2  (rods +  resistors) | 298 | 3.97 | 803 | 7.35 | n/a | n/a |

These CDR BBU calculations use the DARHT-I MOD2 transverse impedance measured with drive rods and resistors installed [22]. This is shown in Figure 6, along with the LAMDA resonance model used for calculating BBU. Because of RF power coupled out of the cavity by the rods, this impedance is significantly less than measured without drive rods, which is the only case that can be calculated with the 2-1/2D AMOS code. Moreover, we use an average of the orthogonal mode impedances that result from the dipole asymmetry introduced by the drive rods [23]. More accurate BBU simulations will be forthcoming after our impedance measurements for a Scorpius test cell.

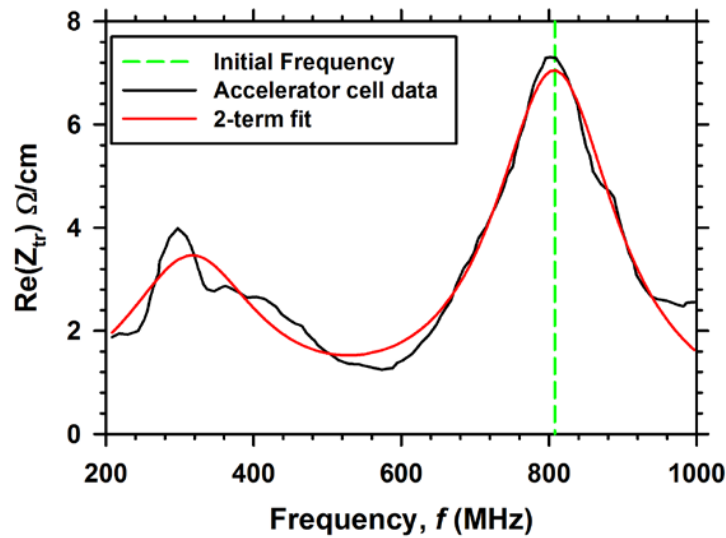

**Figure 6: Measured DARHT MOD2 transverse impedance (black) and LAMDA model (red).**

## *LAMDA BBU Simulations*

Except as noted, LAMDA simulations of BBU with the Scorpius CDR tune used a single frequency initial beam motion to excite the instability, which is analogous to experimental excitation with a tickler cavity, and less ambiguous than excitation from a step offset, or delta function. The frequency chosen for the simulation was the ~808-MHz peak of the dominant mode, as shown in Figure 6. The growth of BBU calculated for Scorpius with the CDR tune is shown in Figure 7. Also shown in this illustration for comparison is BBU growth in DARHT-I calculated by LAMDA with the nominal tune used for many hydrotests. As seen in this plot, the calculated BBU amplification at the exit of Scorpius is slightly less than the calculated amplitude in DARHT.

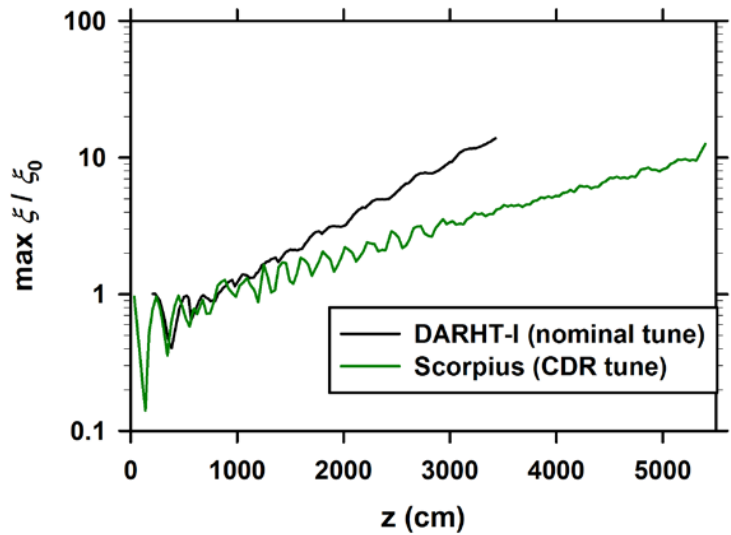

**Figure 7: LAMDA simulation of BBU growth in Scorpius with the CDR tune (green). Also shown is the BBU growth simulated by LAMDA for DARHT-I with its nominal tune (black). As reported in [21], LAMDA simulations of BBU in DARHT-I agree with measurements.**

In practice the instability may be excited by random noise, corkscrew motion, or an offset beam pulse with fast risetime. We simulated the latter with LAMDA. Figure 8 shows the beam displacement from center at the LIA exit resulting from an initial 0.01-cm displacement of the injected beam. As expected, the rapid beam-current risetime excites a wide spectrum which is amplified by the BBU. Figure 9 shows the power spectrum of the BBU at the exit. The two lowest frequency resonances of the transverse impedance model are clearly seen superimposed on the spectrum of the square pulse (compare with Figure 6).

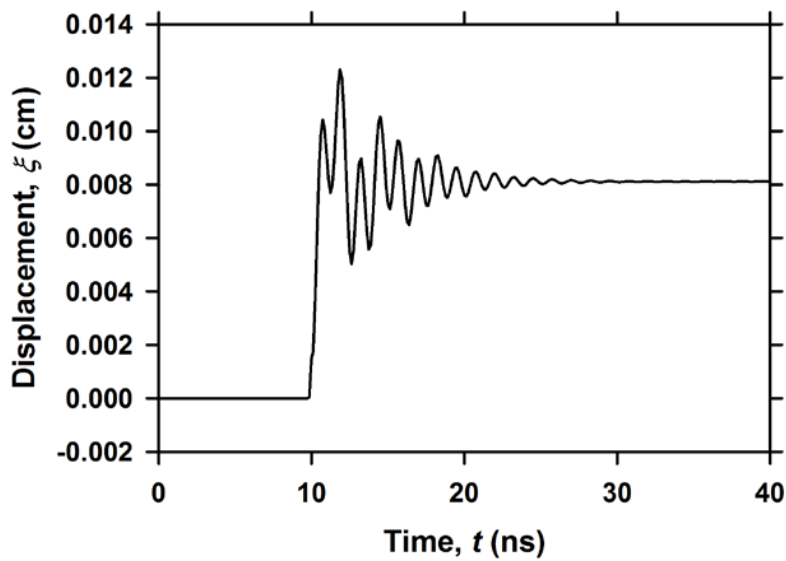

**Figure 8: BBU at LIA exit resulting from injecting a sudden offset pulse into the Scorpius LIA. The initial beam offset for this calculation was 100 microns (0.01 cm).**

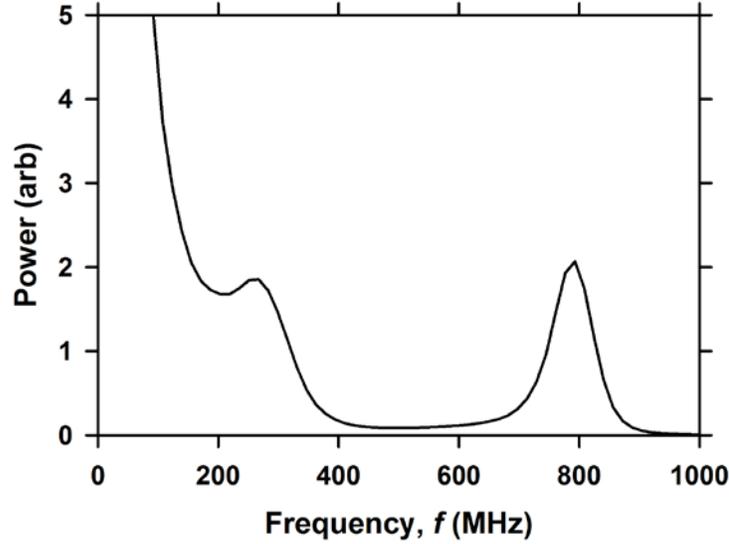

**Figure 9: Power spectrum obtained by Fourier transforming the results shown is Fig. 11.**

These simulations show that BBU growth can be suppressed by increasing the magnetic field of the Scorpius tunes. However, this comes at the cost of more power and cooling for the solenoids. Since the field scales as the magnet current, and the power scales as the square of current, this is a trade that should be carefully considered for any proposed architecture. For the Scorpius CDR architecture, no inter-cellblock magnets are required, and the CDR tune suppresses BBU simulations to less growth than in DARHT-I simulations.

Finally, since LAMDA simulations of BBU in DARHT-I are in agreement with experimental data [21], these results indicate that BBU can be suppressed in Scorpius if the transverse impedance of the Scorpius cell cavity is close to that of the DARHT-I cavity.

## B. Corkscrew Motion

Strictly speaking, corkscrew motion [25] is not an instability. Corkscrew motion results from the interaction of fluctuations of beam electron energy with accidental magnetic dipoles caused by misalignment of the beam transport solenoids. Corkscrew is a concern, because increasing the solenoidal magnetic field to suppress BBU also increases the misalignment fields, thereby increasing corkscrew as a consequence. Although there are means for suppressing corkscrew using corrector dipoles embedded in the LIA cells [26, 27, 28, 29, 14], best engineering practices can do much to mitigate this cause of motion. These practices include stringent alignment tolerances and procedures, and minimization of temporal variation and noise on the accelerating-cell pulsed power.

The amplitude of corkscrew motion can be defined as $A^2 = \left\langle \delta x^2 \right\rangle_t + \left\langle \delta y^2 \right\rangle_t$, where the brackets indicate averaging over time, and $\delta x = x - \left\langle x \right\rangle_t$, $\delta y = y - \left\langle y \right\rangle_t$ [26]. A convenient scaling formula deduced from the analytic theory [25] is given in [30]. This formula predicts that the corkscrew amplitude after $N$ misaligned solenoids would be $A \approx \sqrt{N} \, \phi \, \delta \ell \delta \gamma / \gamma$, where $\gamma$ is the relativistic mass factor, $\delta \gamma$ is its rms variation in time during the pulse, $\delta \ell$ is the rms misalignment, and $\phi$ is the phase advance ($\phi = \int k_\beta \, dz$, where the betatron wavenumber is $k_\beta = 2\pi B_z / \mu_0 I_A$, and $I_A = 17\beta\gamma$ kA ). The cell misalignment includes both tilt and offset, with the tilt contribution approximately equal to the solenoid length times the rms tilt angle, which is added in quadrature to the rms offset.

This scaling has often been used to estimate corkscrew amplitude in presently operational LIAs, as well as to assess new LIA designs [31]. However, it is based on an theory of corkscrew motion of a constant-energy beam coasting through a uniform axial guide field. Since the theory was developed for this highly idealized situation, there is the question of its relevance to accelerated beams in real, non-uniform magnetic fields, such as the Scorpius CDR tune shown in Figure 3. Therefore, we prefer to use beam dynamic simulations to assess Scorpius designs on a case-to-case basis. In particular, LAMDA simulations of corkscrew have found the scaling with phase advance and number of cells to be problematic for large phase advances , such as in Scorpius. This is a result of the detailed dynamics of the beam quasi-cycloidal motion with more than $2\pi$ rotation due to the strong, varying solenoidal field. Moreover, the scaling with misalignment and energy variation is only valid over a narrow range of parameters. Since these are the two parameters of most interest for engineering trade-offs on Scorpius, we focus on a scaling law $A \propto \delta\ell\,\delta V / V$ ; using the cell drive voltage $V$ , rather than the beam energy $\gamma$ , which varies though the LIA.

The accelerating potential fluctuations $\delta V / V$ responsible for corkscrew cover a broad range of possibilities, from completely *incoherent* to *coherent,* which can be defined as follows:

- *Incoherent* – random cell-voltage fluctuations, with no two exactly the same (e.g., random noise).
- *Coherent*- cell voltages have exactly the same variation in time (e.g., identical waveforms).

The actual situation for LIA pulsed power may be intermediate, with both coherent and completely random contributions to the cell voltages. For example, the pulse flattop energy variation on DARHT-I is $\delta\gamma / \gamma < 0.1\%$, and mostly incoherent noise. On the other hand, the variation on DARHT-II is $\delta\gamma / \gamma \sim 2.5\%$, and almost entirely a coherent low-frequency oscillation, which is characteristic of the pulsed power. For equal rms values, coherent cell voltage fluctuations dominate over incoherent, because they add linearly whereas incoherent fluctuations only add in quadrature at best.

For simulations of coherent corkscrew, a cell-voltage waveform was generated by adding normally distributed random voltages with zero mean to a 0.25 MV flattop. This basic waveform was scaled to give rms cell-voltage variations in the range 1% to 10% without changing its fundamental shape. That is, the amplitude of the waveform was changed, but not its shape. This exact waveform was applied to each cell in the simulation. The 1.75-kA flattop current pulse for the simulation was shorter than the cell voltage pulse, so the corkscrew amplitude was only calculated during the current flattop as shown in Figure 10. Incoherent corkscrew was simulated by adding random noise to a nominally 0.25-MV flattop voltage pulse. The random number generator used to define the voltage fluctuations was seeded with a different random number for each cell, thereby ensuring that the variations were truly incoherent.

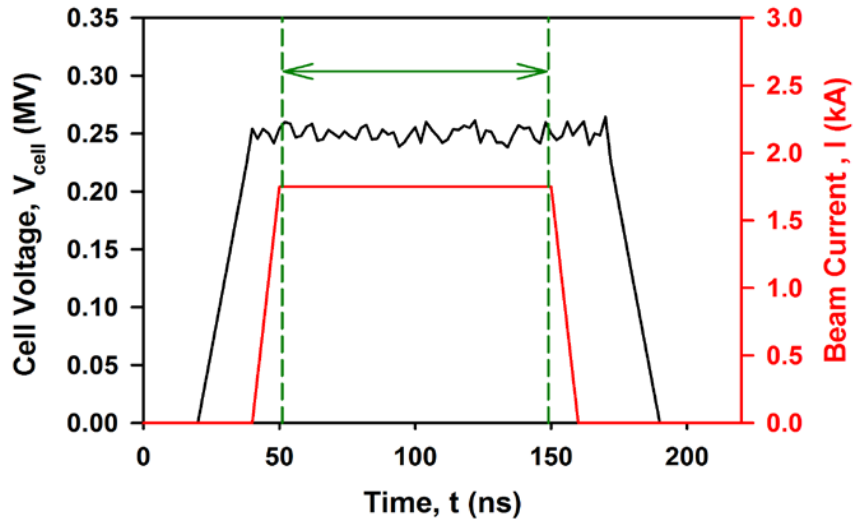

**Figure 10:** .Model cell voltage pulse (black) and beam current pulse (red) showing interval over which LAMDA simulations of corkscrew amplitude was calculated (green). The rms fluctuation on the cell voltage in this illustration is 2.4%.

For reference, measured misalignments on DARHT-II were 0.3-mr rms tilt and 0.1-mm rms offset of the 38-cm long solenoids, giving $\delta\ell$ <0.2 mm. When first assembled, the rms misalignment on DARHT-I was measured to be $\delta\ell \sim 0.2$ mm. therefore, it is plausible to expect similar values for the alignment of Scorpius. Figure 11 shows the transverse magnetic fields calculated by LAMDA for an rms offset $\delta\ell = 0.29$ mm applied to the Scorpius tune of Figure 3. Although this offset is much larger than expected using modern alignment technology, it was chosen for illustrative purposes to accentuate the corkscrew effect.

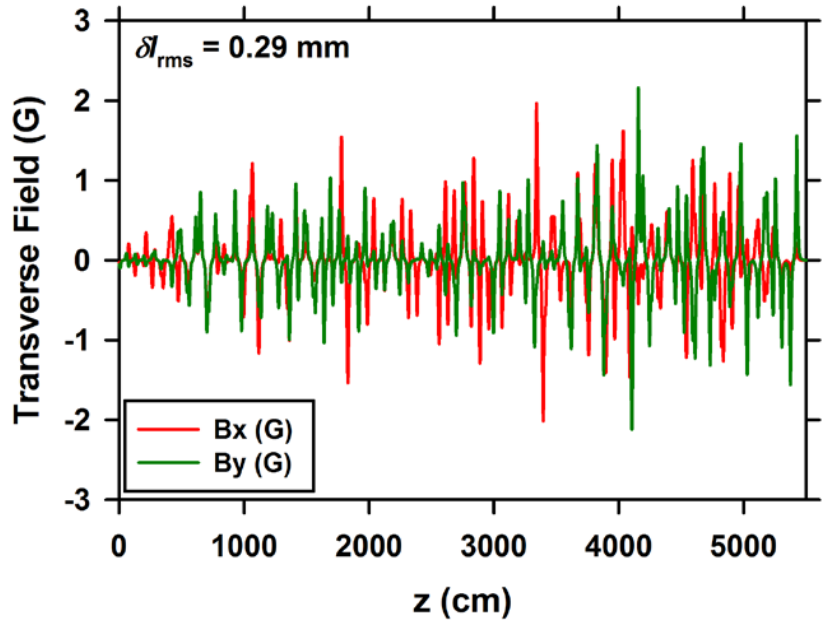

**Figure 11: LAMDA calculation of transverse magnetic fields calculated for the Scorpius solenoidal fields of Figure 3 when misaligned by a DARHT-like rms offset of 0.29 mm.**

Corkscrew amplitudes were calculated from direct simulations by LAMDA. By this we mean that deflections of the beam segments by transverse magnetic fields were calculated directly from the Lorentz force equations. The LAMDA beam dynamics code calculates the beam centroid motion as a function of time [32, 18], including its deflection by the transverse magnetic fields caused by misalignments, like shown in Figure 11. In order to simulate beam motion as a function of time, the beam is articulated by subdividing it into short segments, and applying transverse equations of motion to each segment. Each of these segments has a different energy by acceleration with a time varying cell voltage, like shown in Figure 10.

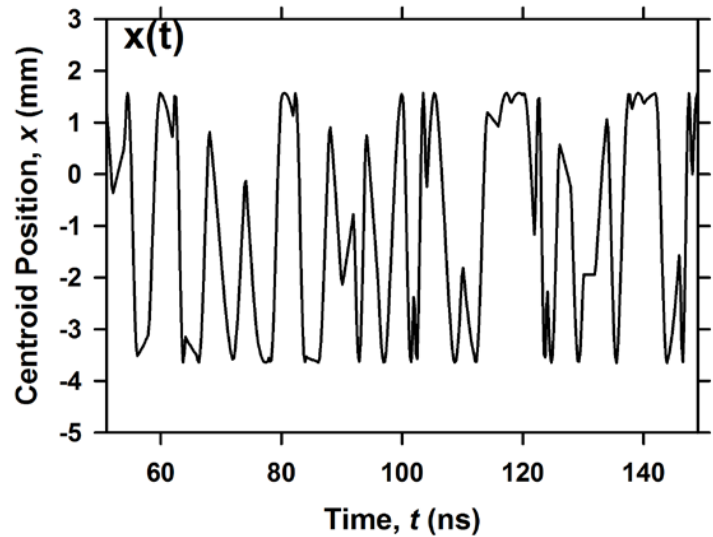

**Figure 12: LAMDA simulation of the y-component of the beam motion at the Scorpius LIA exit. This is the result of applying the voltage waveform with 2.4% fluctuations shown in Figure 10 to every cell, and using the transverse fields shown in Figure 11 calculated for 0.29-mm rms offsets of the solenoids for the tune shown in Figure 3.**

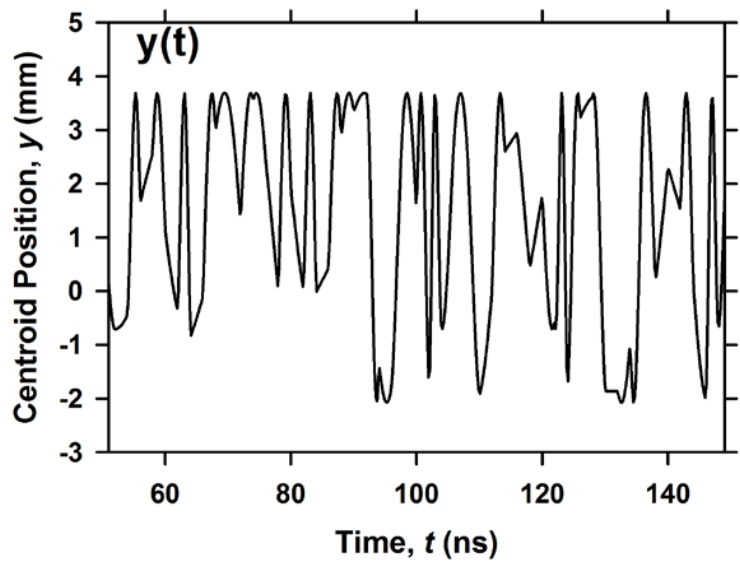

**Figure 13: LAMDA simulation of the x-component of the beam motion at the Scorpius LIA exit. This is the result of applying the voltage waveform with 2.4% fluctuations shown in Figure 10 to every cell, and using the transverse fields shown in Figure 11 calculated for 0.29-mm rms offsets of the solenoids for the tune shown in Figure 3.**

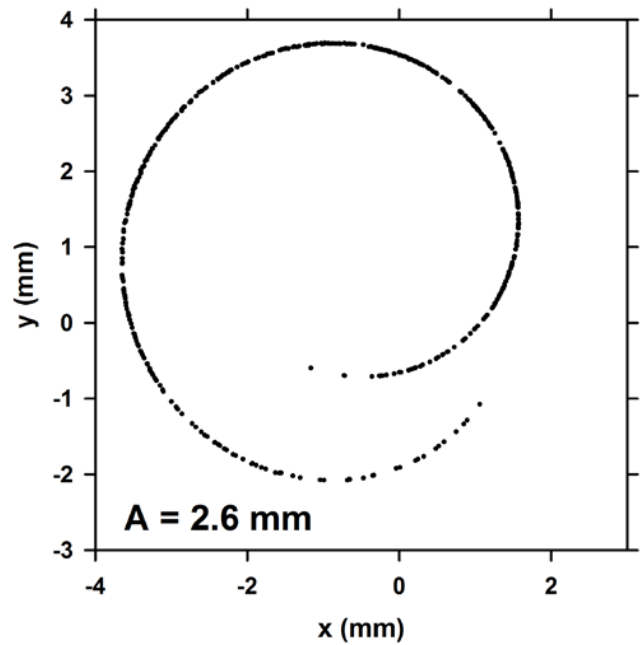

**Figure 14: LAMDA simulation of beam-motion trajectory at the Scorpius LIA exit for the entire time shown in Figure 12 and Figure 13. Individual points are at 1-ns intervals. This is the result of applying the voltage waveform with 2.4% fluctuations shown in Figure 10 to every cell, and using the transverse fields shown in Figure 11 calculated for 0.29-mm rms offsets of the solenoids for the tune shown in Figure 3.**

The resulting beam at the Scorpius LIA exit is shown in Figure 12 and Figure 13 as a function of time. This calculation used the tune, coherent energy variation, and misalignment fields shown in Figure 3 (Scorpius CDR tune), Figure 10 (Coherent 2.4% voltage fluctuations), and Figure 11 (0.29-mm misalignment offsets). Figure 14 shows the trajectory of the motion shown in Figure 12 and Figure 13. The characteristic axial-flux encircling cycloidal trajectory is a dominant characteristic of these corkscrew simulations. The amplitude calculated for this motion at the LIA exit (z = 5350 cm) was $A$ = 2.6 mm.

This cycloidal motion was also prominent in early experimental data from DARHT-II, and was tuned out using corrector dipoles in the cells [29, 14]. For suppressing corkscrew with corrector dipoles, we actually use a more stringent measure of amplitude as a metric; the diagonal $D$ of the rectangular area bounding the trajectory [28]. The reason is that $D$ bounds the spot-to-spot wandering for four pulses taken during different times on the curve. For the example shown in Figure 14 this metric would be $D \sim$ 7.9 mm which is much greater than $A$. It is noteworthy that on DARHT-II we were able to use the corrector dipoles to reduce $D$ from $D \sim$ 1 cm to less than 2 mm [14, 17], so mitigation of the simulated Scorpius corkscrew is well within the demonstrated capability of this technique.

Two fundamental properties of the corkscrew cycloidal motion at the LIA exit are:

1)    The angular position depends on beam energy, so the total angle subtended in an y(x) plot like Figure 14 is proportional to the rms variation in cell voltage [29]. This angle can exceed $2\pi$ for large fluctuations (multiple loops) and large phase advance, and then there is a saturation effect, with the size metrics $A$ and $D$ no longer depending on $dV/V$.

2)    The radial size of the cycloidal trajectory depends on the misalignment rms offset. Therefore, no saturation effect is evident in misalignment offsets.

These characteristics are clearly evident in the LAMDA simulations.

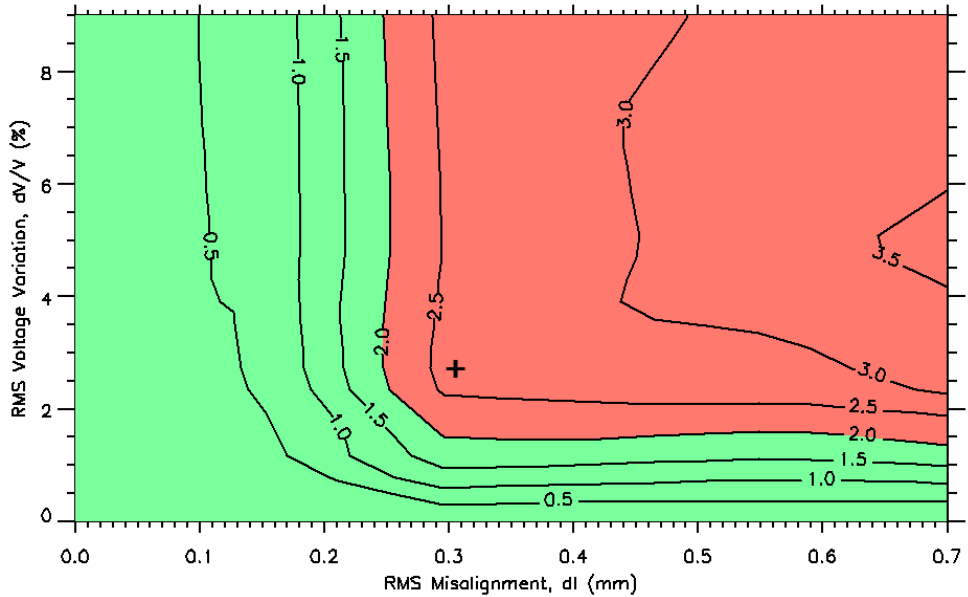

**Figure 15: Contours of equal corkscrew amplitude *A* as a function of misalignment offset dl and cell voltage variation for the Scorpius CDR tune. The green shading indicates the region for which the amplitude would be less than ~ 2 mm. The single point plotted as a cross indicates the offset and pulsed-power fluctuation for the simulation illustrated by Figure 10 through Figure 14.**

To clarify the engineering trade space of misalignment and pulsed-power quality, we performed numerous LAMDA simulations over a wide range of misalignment and voltage fluctuation parameters. The results of these are plotted as contours of corkscrew amplitude as a function of both misalignment and cell-voltage fluctuation (Figure 15). The green shading indicates the trade space for which the LAMDA simulations produce less than 2-mm corkscrew amplitude. On this graph, a linear scaling of amplitude with the product of misalignment and voltage fluctuation, $A \propto \delta\ell \, \delta V / V$, would generate hyperbolic contour curves. This is evidently only the case in the lower left-hand corner, $\delta V / V < \sim 2.5\%$ and $\delta\ell < \sim 0.3$ mm. This is apparently the limiting boundary of the idealized theory for this tune of Scorpius.

This plot also shows that for larger pulsed-power fluctuations the amplitude saturates. This is because the rotational angle subtended by the quasi-cycloidal motion of the beam position is greater than $2\pi$, and further increases in $\delta V / V$ do not increase the overall size of y(x) plots, such as Figure 14. Thus, if modern alignment practices are adhered to, substantial fluctuations on the cell pulsed-power can be tolerated.

## C. Image Displacement Instability

The image displacement instability (IDI) is also the result of a slightly offset beam interacting with a cavity [33, 34, 35, 36]. While the BBU is the result of beam deflection by specific cavity resonances, the IDI has no frequency dependence. It can be derived from the beam deflection due to static fields. Therefore, it can disrupt the beam even at the lowest frequencies. Moreover, unlike the BBU, the IDI has a definite stability threshold. That is, the beam is unstable in a guide field less than $B_{\min}(\gamma, I_b)$, which is a function of beam energy, current, and accelerator geometry. Thus, it is most dangerous at the entrance to the accelerator, where the magnetic field is low (see Figure 3).

The IDI is the result of the magnetic and electric field boundary conditions, which can be satisfied by an image current and image line charge. A beam slightly offset from the center of a beam pipe is attracted to the wall by the image of its space charge, and repelled from the wall by the image of its current. These forces balance to within $1/\gamma^2$, with the net force being attractive toward the wall. This is normally counterbalanced by the focusing field. However, in the vicinity of a gap in the wall, the induced charge on the wall collects at the gap edges, and the electrical attraction is almost unchanged. That is,, if the gap is short compared to the tube radius, the position of the image line charge is almost unchanged. On the other hand, the azimuthal magnetic field of the beam decays with radius in the cavity exactly as in a pipe with radius equal to the outer wall of the cavity, and the effect is as if the current mage is located at a greater distance, reducing the repulsive force from the wall. Thus, each cavity presents an additional deflecting force toward the wall that must be overcome by the focusing force. Therefore, periodically spaced cavities can be modeled as a periodic modulation of the restoring force in the equation of motion for the beam centroid. This suggests parametric amplification of the displacement; indeed, after suitable coordinate transformations, the equation of motion can be written as the Mathieu equation [35], which is a well-known model for such instabilities. In canonical form, the Mathieu equation is

$$\frac{d^2\psi}{d\zeta^2} + \left(a - 2q\cos 2\zeta\right)\psi = 0 \ , \tag{5}$$

with $\varsigma = 2\pi z / L$ , and

$$
\begin{aligned}
a &= \left(\frac{k_\beta L}{\pi}\right)^2 - \left(\frac{2I}{I_A\beta}\right)\left(\frac{wL}{\pi^2 b^2}\right) \\
q &= 2\left(\frac{2I}{I_A\beta}\right)\left(\frac{wL}{\pi^2 b^2}\right)\frac{\sin\vartheta}{\vartheta} \quad , \\
\vartheta &= \frac{\pi w}{L}
\end{aligned}
\tag{6}
$$

where $w$ is the gap width, $b$ is the tube radius, $L$ is the inter-gap spacing. The boundaries of stable solutions are well known functions $a_n(q)$, and are shown in Figure 16.

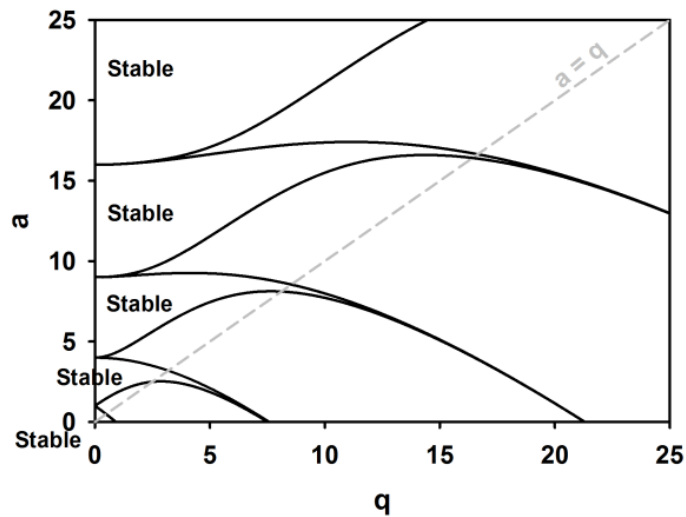

**Figure 16: Regions of stable and unstable solutions to the Mathieu equation.**

In particular, for parameters relevant to Scorpius, $q << 1$ and stability obtains for $0 \leq a \leq 1 - q - q^2 / 8$, which gives the minimum magnetic guide field for stability

$$B_{kG}^2 > 1.36\gamma \left(\frac{w}{L}\right)\frac{I_{kA}}{b_{cm}^2} \ . \tag{7}$$

Another theory approximating the disturbance as a wake field effect also finds the same minimum field for stability scaling, but with a somewhat larger constant of proportionality [22]. Both theories predict stability for the average Scorpius magnetic field in the first block of cells. Table III shows the minimum field required to stabilize the IDI compared with the average field in the first 8-cell block of Scorpius. Since the average Scorpius field is almost twice the most conservative stabilizing field, the IDI is not expected to be a problem.

**Table 3. Threshold Field for the Image Displacement Instability.**

|  | Scorpius |  | Briggs[22] | Caporaso and Chen[21] |
|---|---|---|---|---|
| $\langle KE \rangle$ | $I$ | $\langle B \rangle$ | $B_{min}$ | $B_{min}$ |
| MeV | kA | G | G | G |
| 3.1 | 2.1 | 333 | 176 | 101 |

## D. Diocotron Instability

The Scorpius diode might produce a hollow beam. Therefore, we assess the possibility of diocotron instability, because hollow beams in axial magnetic fields can be unstable under some conditions [37, 38]. The theory of this instability is well founded and has been validated by numerous experiments with both neutral and non-neutral plasmas and relativistic electron beams. Diocotron would be a troublesome source of beam emittance if present on the Scorpius beam under normal operating conditions.

The diocotron is an interchange type of instability caused by sheared rotational velocity in a beam with a radial density profile having an off-axis maximum, as in a concave beam ("inverted") profile. Since the diocotron is driven by sheared flow in a medium with a density gradient, it is analogous to the Kelvin-Helmholtz (KH) instability in fluids. In a uniform axial magnetic field, the rotational shear is due to the $\mathbf{E} \times \mathbf{B}$ drift produced by beam space charge. Since the electric field is due to the beam space charge, this driving force is reduced by

The instability is characterized by a strength parameter

$$s = q = \omega_p^2 / \omega_c^2 \tag{8}$$

where $\omega_p^2 = e^2 n_e / \gamma m_e \varepsilon_0$ and $\omega_c = eB / \gamma m_e$. Thus, $s = \gamma n_e m_e / \varepsilon_0 B^2$, and characterizes the ratio of space charge force driving the instability to stabilizing magnetic focusing force. Beams with greater $s$ are more unstable; that is, high density beams in weak magnetic fields are most unstable.

Diocotron theory has largely been developed through numerical solution of dispersion relations for special geometrical cases. These have shown that the growth rate is proportional to $\omega_D / \gamma^2$, where the diocotron frequency is $\omega_D \equiv \omega_p^2 / 2\omega_c$. The factor of $1 / \gamma^2$ in the growth rate is a reflection of the depression of the $\mathbf{E} \times \mathbf{B}$ drift velocity by the same factor when self-focusing by the beam current is accounted for. The geometrical constant of proportionality has a maximum value near unity for the extreme case of a thin annular beam. Thus, the maximum growth rate is $\omega_D / \gamma^2$, and the growth rate for the Scorpius solid beam is expected to be much less.

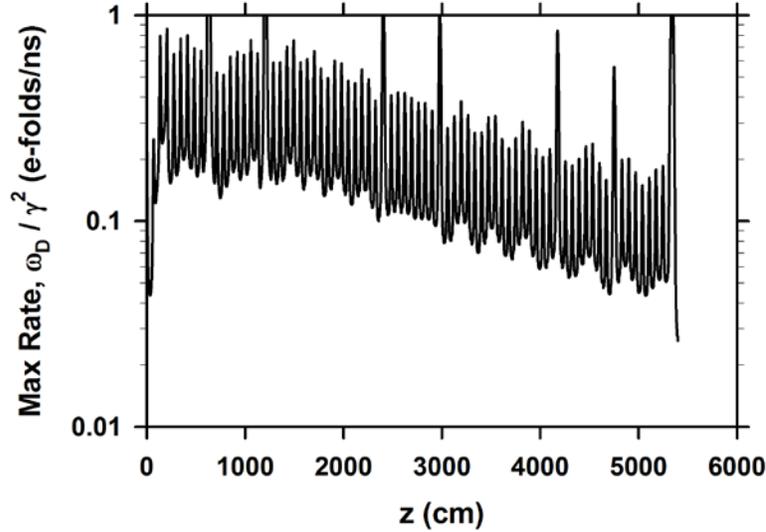

**Figure 17: Maximum diocotron growth rate $\omega_D / \gamma^2$ for the Scorpius CDR tune. Since this rate is for a thin annular beam, the Scorpius rate is expected to be significantly less. This plot clearly shows the improvement in stability achieved by increased beam energy and increased magnetic guide field.**

The maximum rate for a thin annular beam in the Scorpius tune is shown in Figure 17 . According to this plot, the geometrical factor describing the beam profile would have to be less than ~0.1 to prevent more than a couple of diocotron e-folds during the beam pulse (growth rate < .02 efolds/ns). The best means for predicting the profile, and resulting growth, is through PIC simulations of the diode, injector, and LIA. Finally, the best strategy for Scorpius may be to prevent the diocotron by designing a diode that produces a convex profile, which has been shown to be absolutely stable.

## E. Resistive Wall Instability

The resistive wall instability is usually thought to be a concern only for long-pulse relativistic electron beams [39, 40, 41, 42]. However, with the advent of multi-pulse, high-current linear induction accelerators (LIAs) [43, 44, 45, 46, 47] the possibility of pulse-to-pulse coupling of this instability has been demonstrated through simulations [48]. In earlier papers [49, 50] estimates of instability growth in drift transport regions of multi-pulse machines were based on analytic theory. A quantitative examination of the pulse-to-pulse coupling phenomenon using direct simulation of the forces on the beam is the thrust of this secction.

The instability results from the Lorentz force on the beam due to the beam images charge and current in the conducting beam pipe. If the pipe is perfectly conducting, the electric force due to the image charge attracts the beam to the pipe wall. However, the magnetic force due to the image current repels the beam from the wall. For a relativistic beam, these forces almost cancel, leaving an attractive force equal to $1 / \gamma^2$ times the image charge force, where $\gamma$ is the Lorentz relativistic mass factor. However, if the beam pipe is not perfectly conducting, the magnetic field due to the image current decays on a time scale $\tau_d \propto \sigma$, where $\sigma$ is the pipe conductivity. Thus, if the beam pulse length $\tau_p$ is greater than $\tau_d$ the magnetic repulsion of the beam tail will be weaker than the repulsion of the beam head. In the absence of an external focusing force, this causes a head-to-tail sweep of the beam toward the wall. This sweep grows as the beam propagates down the pipe. The strong external focusing force provided by the solenoidal magnetic fields of an LIA complicate this simple picture, but generally do provide significant suppression of instability growth.

For a constant-current coasting beam the time-varying electromagnetic fields produced by the conducting-wall images of a beam displaced a distance $\xi(t)$ from the centerline were derived in [40]. These fields are proportional to $\xi(t)$, and for early times, the resulting radial force on the beam toward the wall is approximately

$$\mathbf{F}(z,t) = \mathbf{e_x}\left[\frac{2eI}{b^2\beta c\gamma^2}\xi(z,t) + \frac{4eI\beta}{\gamma\pi b^3\sqrt{\sigma}}\int_{-\infty}^{t}\frac{d\xi}{dt'}\sqrt{t-t'}dt'\right] \tag{9}$$

where $e$ is the electron charge, $I$ is the beam current, $b$ is the beam pipe radius, $\beta\gamma = \sqrt{\gamma^2-1}$ is the normalized electron momentum [41]. Except where noted, cgs units are used in this section. The first term is the force that would be applied for a perfectly conducting wall, and the integral term represents the decay of the magnetic repulsion due to diffusion of the dipole field.

Eq. (9) shows that the transverse force on every beam electron is due to the conducting tube wall images of the preceding electrons, but not of those following it (due to relativistic causality) [51]. Thus, the transverse force on the last electron in the pulse is due to the images of all of the rest of the electrons preceding it in the pulse. It follows that an electron in a second pulse following the first is also subject to a transverse forces due to images of all of the electrons preceding it, including those in the first pulse. This is pulse-to-pulse coupling. The simplest example is that if there is no gap between the two pulses, this case is indistinguishable from a single pulse of greater length than the first.

A key result of the analytic theory is that the characteristic distance for significant growth is

$$z_g(\tau) = \frac{4\gamma}{I_{kA}\sqrt{\tau_{\mu s}}}\left(\frac{\sigma_{s^{-1}}}{10^{17}}\right)^{1/2}\left(\frac{b_{cm}}{10}\right)^3\left(\frac{100}{2\pi}k_{0m^{-1}}\right) \quad \text{m} \tag{10}$$

where $\tau$ is time into the pulse measured back from the head, and the units for the parameters are indicated by the subscripts [41]. Here, the space-charge reduced betatron wavelength $k_0$ is defined by

$$k_0^2 = k_\beta^2 - \frac{2I_{kA}}{17(\beta\gamma)^3}\frac{1}{b^2} \tag{11}$$

Worth noting is the strong dependence of the characteristic distance on the size of the beam pipe; $z_g \propto b^3$, which favors large beam pipes for long-pulse machines. Simulations with the LAMDA beam-dynamics code follow this scaling law [48].

We used LAMDA [18] to investigate this instability for the Scorpius design using the CDR tune. For these simulations we used the 4-pulse format shown in Figure 17 for the injected beam. Since the worst case for this instability is with the longest pulses having the least inter-pulse dwell time, we simulated pulses with a 80-ns flattop, 10-ns rise and fall, and inter-pulse separation of 200 ns. Figure 18 shows LAMDA calculations of the resistive wall instability amplitude at the end of the pulse flattops, where growth is maximum for each pulse. There is no evidence of growth or coupling. The magnetic focusing field of Figure 3 is strong enough that the instability is damped. Thus, pulse-to-pulse coupling is not expected to be a problem in the Scorpius LIA. However, coupling might be a problem in long drift sections of the downstream transport, unless higher conductivity or larger size beam pipes are used there.

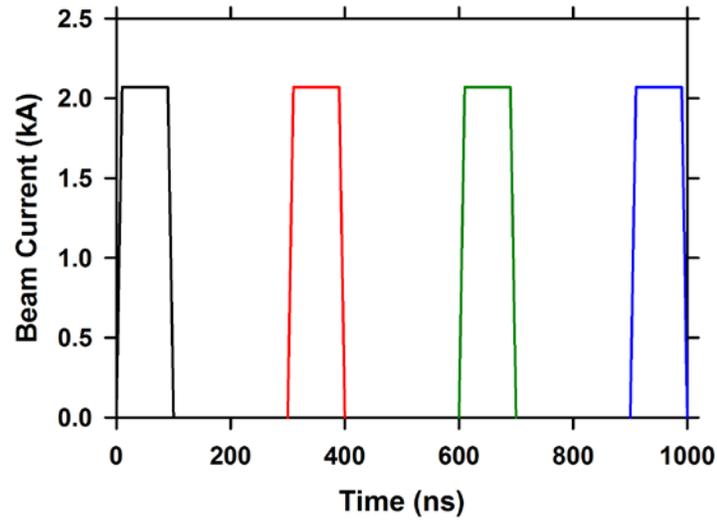

Figure 18: Four-pulse format of the simulated electron beam injected into the Scorpius tune shown in Figure 3. Each pulse have 80-ns flattop and 10-ns rise and fall times The inter-pulse separation is 200 ns. Instability growth was calculated at the end of the flattop for each pulse.

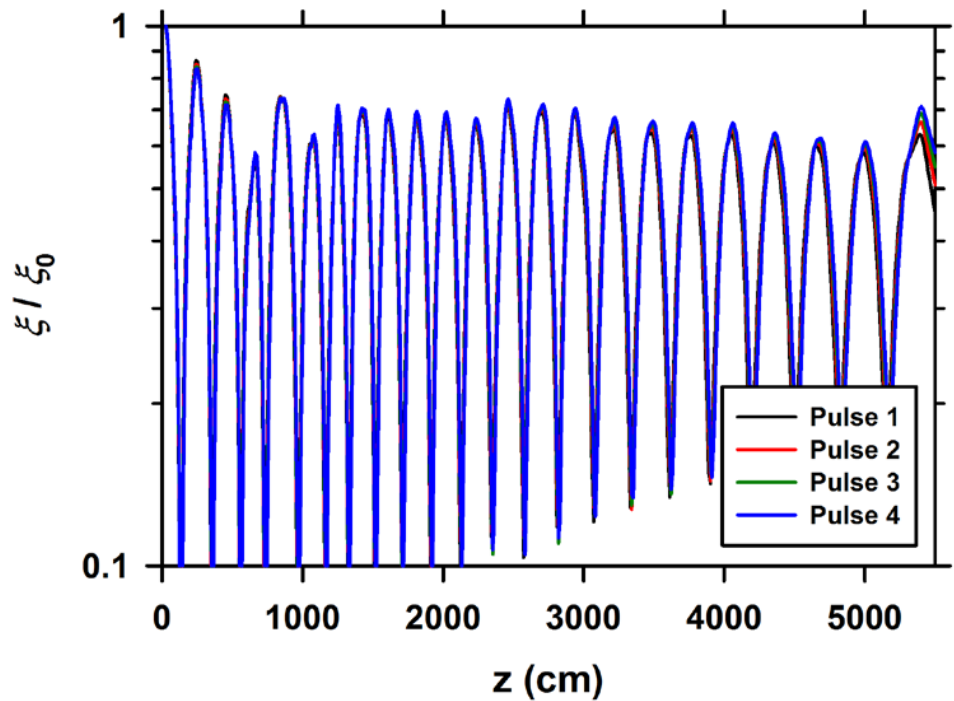

Figure 19: LAMDA simulation of the instability attenuation due to the Scorpius magnetic focusing shown in Figure 3. (black) Tail of pulse #1. (red) Tail of pulse #2. (green) Tail of pulse #3. (blue) Tail of pulse #4.

## F. Ion Hose Instability

Another instability that can be dangerous for a high-current accelerator is the ion-hose instability [52]. This is caused by beam-electron ionization of residual background gas. The space-charge of the high-energy beam ejects low-energy electrons from the ionized channel, leaving a positive channel that attracts the beam. This causes the beam to oscillate about the channel position. Likewise, the electron beam attracts the ions, causing them to oscillate about the beam position. Because of the vast differences in particle mass the electron and ion oscillations are out of phase, and the oscillation amplitudes grow. The fastest growing frequency is predicted to be

$$\omega_{i0} = 1.22 \frac{c}{R_{rms}} \sqrt{\frac{m_e I}{m_i I_0}} \tag{12}$$

where the factor of 1.22 comes from the detailed theory [53], $m_e$ and $m_i$ are the electron and ion masses, $R_{rms}$ is the rms beam radius, $I$ is the beam current, and $I_0 = 17.045$ kA   Typical beam parameters and background gases result in very low frequencies (of tens of MHz).

This instability was of some concern for the long-pulse DARHT-II LIA, and a substantial effort was devoted to understanding this instability through theory and experiments. Like the BBU, this instability is convective, with oscillations growing in time at a fixed location in the accelerator, and with the maximum amplitude of these in the beam pulse growing with distance through the accelerator. The analytic theory [53] predicts the growth of a perturbation $\xi(z,t)$ to be given by $\max \xi / \xi_0 = \exp \Gamma(z)$ in the so-called linear regime, with

$$\Gamma(z) = 4.75 z \frac{f \nu m_e c^2}{e B R_{rms}^2} \quad \text{(cgs)} \tag{13}$$

where $f$ is the fractional beam neutralization by the channel ions, and $\nu = I / (m_e c^3 / e) = I_{kA} / 17.045$. Neglecting recombination, impact ionization of the residual background gas is governed by the rate equation $dn_i / dt = (n_{n0} - n_i) n_e \langle \sigma v_e \rangle$, with $n_e \langle \sigma v_e \rangle \leq 1/\text{ms}$ for typical beam and gas parameters. Therefore, the fractional ionization for pulselengths less than ~1 ms is roughly given by $f = pt / \alpha$, where $p$ is the background pressure, and $\alpha$ is called the ionization constant, usually expressed in Torr-ns. The dominant residual gas species in DARHT has been found to be water, with $\alpha = 1.11$ Torr-ns. Putting all this into practical units gives

$$\Gamma_m = 0.043 I_{kA} \tau_{\mu s} L_m \left\langle p_{\mu Torr} / \left( B_{kG} R_{cm}^2 \right) \right\rangle , \tag{14}$$

where the brackets denote averaging over the LIA length $L_m$. In this equation, the product of current, pressure and pulse length is proportional to the channel-ion density, which shows the fundamental property of growth depending directly on the attractive force between electrons and ions.  We experimentally confirmed Eq. (14) scaling on DARHT-II over a wide range of beam and gas parameters [15]. Since $\Gamma$ directly depends on the neutralization fraction, which is negligible for typical short-pulse LIA vacuum and beam parameters, the ion-hose instability has only been a concern for long pulse accelerators in the past.

On the other hand, ion hose is a concern for multiple-pulse LIAs like Scorpius, because recombination of the ion channel is an exceedingly slow process, taking place over time scales much longer than the inter-pulse separation. Thus, although the first pulse may not be long enough to provide sufficient ionization for growth, the ion density at the time of the fourth pulse may be great enough to cause problems according to Eq. (14). The predominant recombination processes for impact ionized residual background gas are radiative and three body. For cold channels the collisional rate dominates. The collisional recombination rate is given by [54]

$$R_c = 1.4 \times 10^{-31} n_i n_{cl}^6 \left( \chi / kT \right)^2 \exp\left[ \chi / \left( n_{cl} + 1 \right)^2 kT \right] \quad (\text{cm}^3/\text{s})$$

$$n_{cl} = 126 Z^{14/17} n_i^{-2/17} \left( \chi / kT \right)^{-1/17} \exp\left[ 4\chi / 17 n_{cl}^3 kT \right]$$

(15)

where $\chi$ is the ionization energy and $n_{cl}$ is the quantum number of a collisional limit. For example, a fully ionized channel in 1.0-μTorr residual background has a collisional recombination time of ~32 μs at room temperature. This is a minimum, the recombination time increases for lower density and/or higher temperature. Therefore, it is safe to neglect recombination between pulses in a Scorpius pulse train < 3 μs duration.

In order to determine the Scorpius vacuum requirements for preventing ion hose we performed LAMDA simulations with all four pulses stacked end to end, which is equivalent to neglecting recombination during the dwell time between well-separated pulses. The total pulse-length was equivalent to four 80-ns current pulses. We used the CDR tune shown in Fig. 8 for these simulations, but we used a constant beam radius as we did for earlier simulations of DARHT-II [53, 55]. Other parameters are the same as used for other Scorpius simulations, as shown in Table V.

**Table 4. Parameters used in LAMDA ion hose simulations**

| Parameter | Symbol | Units | Value |
|-----------|--------|-------|-------|
| Initial Kinetic Energy | $KE_0$ | MeV | 2.10 |
| Increment/cell | $\delta KE$ | MeV | 0.25 |
| Beam Current | $I$ | kA | 2.07 |
| Beam RMS Radius | $R_{rms}$ | cm | 0.5 |
| Pulse Flattop | $\tau$ | ns | 320 |
| Pulse Rise/Fall | | ns | 10 |
| Residual Gas | $H_2O$ | | water |
| Neutralization Constant | $\alpha$ | Torr-ns | 1.11 |
| Resonant Frequency | $f_0$ | MHz | 22.3 |

Simulations to determine growth for different background pressures used a single frequency initial excitation at the resonant frequency predicted by Eq.(12), $f_0 = 22.3\,\text{MHz}$. Results of growth for three pressure (100 nTorr, 400 nTorr, and 1 μTorr) are shown in Figure 19. From these simulations the threshold for amplification less than a factor of two is ~150 nTorr. Thus, interlocking the Scorpius vacuum system should provide an adequate margin of safety for the ion-hose instability.

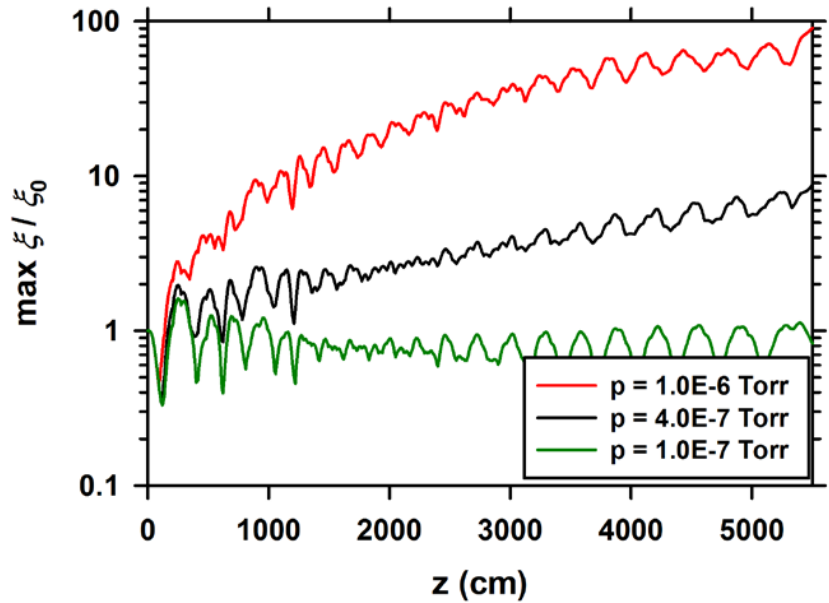

**Figure 20: Ion hose instability growth in Scorpius CDR tune for different residual background pressures of H2O. Red curve: Residual pressure 1.0E-6 Torr. Black curve: Residual H2O pressure 4.0E.7 Torr. Green curve: Residual H2O pressure 1.0E-7.**

We also examined growth excited by a rapidly rising beam offset by starting the offset at the beginning of the current flattop. The resulting ion hose at the exit of the LIA is shown in Figure 20. The power spectrum of this waveform (Figure 21) clearly shows the dominant frequency to be the resonance predicted by theory (Eq.(12)).

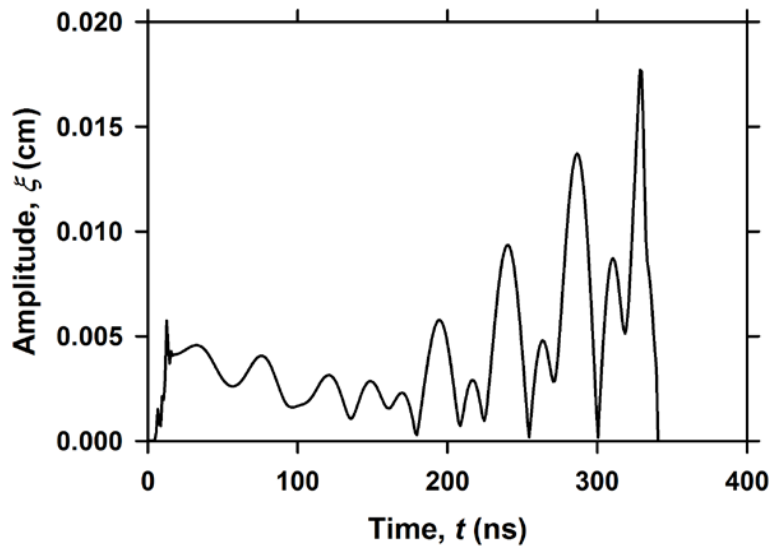

**Figure 21: Ion hose at the LIA exit excited by a step offset of x = 0.707E-3 cm starting at 10 ns. This offset is 1% of the beam envelope radius. The background pressure for this simulation was 1.0 µTorr.**

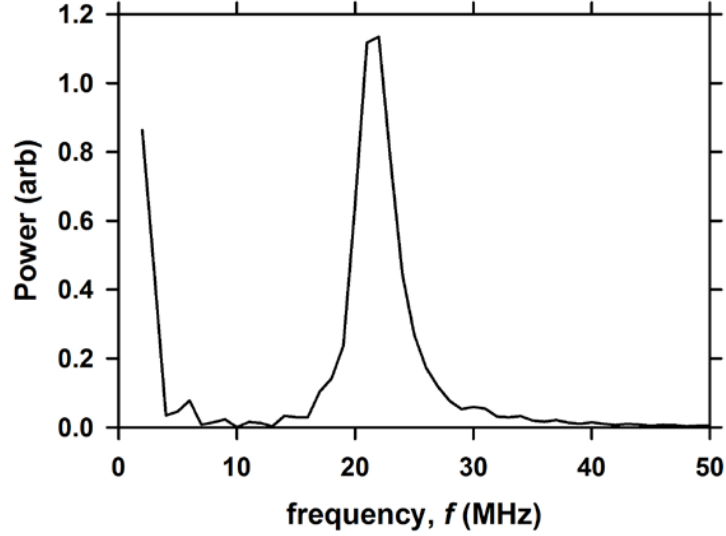

## G. Parametric Envelope Instability

As seen in Figure 3, the Scorpius magnetic focusing field is periodically modulated. Moreover, the envelope of a slightly mismatched beam undergoes m=0, "breathing mode" oscillations (see Figure 5 for example). Under some circumstances, beam transport in a spatially modulated magnetic field can cause a parametric instability of these envelope oscillations [56], which in turn could cause halo and emittance growth [14].

This instability can be explored by considering the envelope equation for a beam coasting through a constant magnetic field [31, 32];

$$\frac{d^2 r}{dz^2} = -k_\beta{}^2 r + \frac{K}{r} + \frac{\varepsilon^2}{r^3} \ . \tag{16}$$

Here, $r$ is the radius of the equivalent uniform beam, which is related to the rms radius of the actual distribution by $r = \sqrt{2} r_{rms}$. Also, $K = \left(2/\beta^2\gamma^2\right)\left(I_b/I_A\right)$ is the generalized perveance and $\varepsilon$ is the beam emittance. For a given beam energy, current, and emittance, a constant envelope radius can be found by setting the right hand side of Eq. (16) to zero describing a matched beam with constant envelope radius;

$$r_m^2 = \frac{1}{2k_\beta^2}\left[ K + \sqrt{K^2 + 4k_\beta^2\varepsilon^2} \right] \ . \tag{17}$$

Now, by solving the envelope equation for small perturbations on this matched radius, the wavenumber of these oscillations is found to be

$$k^2 = k_\beta^2 + \frac{K}{r_m^2} + 3\frac{\varepsilon^2}{r_m^4} \quad . \tag{18}$$

In a uniform field magnetic field, these are stable, but if the focusing field is periodically modulated, they may be parametrically amplified, especially if the field modulation is in resonance with the natural wavelength. For focusing-field is modulation with wavelength $L$ (e.g. cell length or magnet spacing), the equation for the envelope perturbations can be reduced to a Mathieu equation with parameters

$$a = \left(\frac{kL}{\pi}\right)^2$$
$$q = \left(\frac{k_\beta L}{\pi}\right)^2 \frac{\delta B}{B} \tag{19}$$

which has well-known parametric regions of instability (Figure 16). Fortunately, Eq. (19) indicates that $a > q$ always, and Figure 16 shows that region to have small zones of instability. For full-energy Scorpius beam parameters (20-MeV) it is unlikely that a highly-modulated magnetic tune would cause instability.

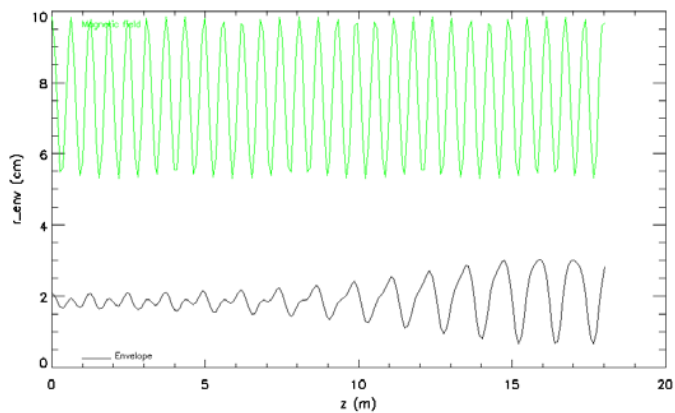

**Figure 23: Unstable envelope oscillations of a low-energy beam (1.5-MeV, 2-kA) coasting in a periodic guide field (450G average with 30% modulation) as calculated with the envelope equation. (black) Beam envelope. (green) Solenoidal focusing field.**

This instability is most troublesome for low energy beams, so it might be a problem if Scorpius is operated in a mode for low-dose radiography of thin objects. In this case, only a few cells might be used to accelerate the beam, allowing it to decelerate through the rest of the inactive cells to a low enough endpoint energy for low-dose radiography. For example, Figure 22 shows the solution to Eq. (16) for a 1.5-MeV, 2-kA beam coasting through a 450-G average guide field with 30% modulation with $L$=0.62 m (close to the Scorpius cell spacing). With these parameters, the beam is clearly unstable.

The signature of this instability, as seen in Figure 22, is growth of initially small envelope oscillations. In order to test the Scorpius CDR tune for resilience to this effect we initialized a slightly mismatched beam in our XTR envelope code. Very little, if any, growth of initial perturbations was observed (Figure 23), demonstrating that this envelope instability should not be a problem for Scorpius.

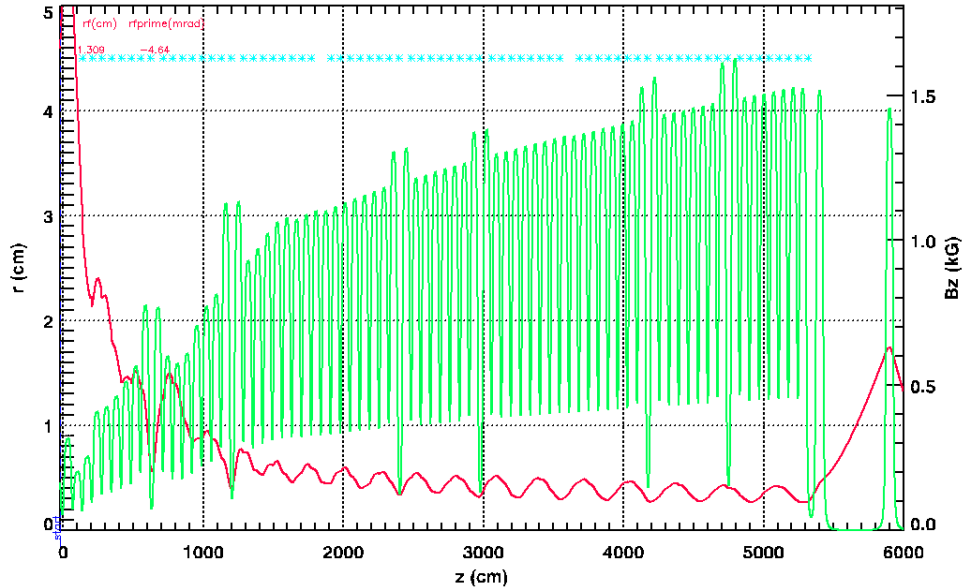

# V. Mitigation Measures

We have engineered several features into Scorpius to mitigate the instabilities discussed above. We summarize them here. Details of each are found in detailed engineering documents.

- BBU; DARHT-I cavity transverse impedance, strong magnetic field
- Corkscrew; corrector dipoles in cells, accurate alignment, flat accelerating voltages, strong guid field
- IDI; strong magnetic field
  - Since the average Scorpius field in the first cell block is almost twice the most conservative stabilizing field, the IDI is not expected to be a problem.
- Diocotron; convex beam profile, high injected beam energy, strong magnetic field
- Resistive Wall; high conductivity beam pipes in drift regions, strong magnetic fields
- Ion Hose; Distributed pumping for ultra-low vacuum, strong magnetic field

It is no surprise that a strong magnetic guide field provides a restoring force that suppresses most of these. Therefore, robust magnetic magnets and power supplies are critical for the success of Scorpius.

# VI. Accelerator Beam Diagnostics

## A. Invasive Diagnostics

### *Beam Imaging*

Scorpius will incorporate the capability for imaging the beam at several locations; at the LIA entrance, one-third of the way through the LIA, two-thirds of the way through the LIA, and at the LIA exit. These positions are occupied by transport solenoids that can be easily removed and replaced with an imaging station that incorporates an insertable imaging target.

Streak and framing cameras will record images of beam generated optical transition radiation (OTR) light from targets inserted in the beam line at 45° to the axis. The OTR targets will be thin titanium foils. We have found these to be the most robust of all the materials we have experimented with, although fused silica targets emitting Cerenkov radiation produce significantly more light. Target insertion and retraction will be remotely controlled from the control room in order to expedite operations. Several camera options are possible for recording the beam images. These can be broadly categorized as either framing or streak cameras.

Beam imaging is at the heart of the technique used to find the beam emittance and beam parameters at an inaccessible location, such as the exit of the diode. The technique is to use a single solenoid to focus the beam on an imaging target. Beam images are recorded for a succession of shots with varying magnet focusing strength. A beam size is deduced from the images for each shot. The usual beam-size metric is $y_{rms}$, which is the rms width of the line spread function (projection), because it is insensitive to errors in the target rotation angle. For an azimuthally symmetric beam current profile it can be shown that $R_{rms} = 2 y_{rms}$, where $R_{rms}$ is the beam image rms radius. An envelope code is then run multiple times to find the initial values of beam radius, convergence angle and emittance that give the best fit to the data.

## Framing Cameras

Framing cameras take snapshots of the image. The DARHT single-frame Princeton Instruments PiMax cameras use gated micro-channel plate (MCP) image intensifiers coupled to CCD arrays with fiber-optics. These single frame cameras feature high resolution (1024 x 1024 pixels) and gating as fast as 2 ns. An example image of a kicked DARHT-II pulse in OTR light from a titanium foil is shown in Figure 24. The spatial detail of the beam current distribution that is available with this system is evident.

Multiple PiMax cameras can be combined with optical beam splitters to capture multiple images of a single beam pulse. Purpose-built beam-splitter based, multiple-frame cameras (e.g., Invisible Vision UHSi12/24) can provide up to 24 frames with 5-ns or greater inter-frame times, which would provide ~10 images during a single Scorpius pulse.

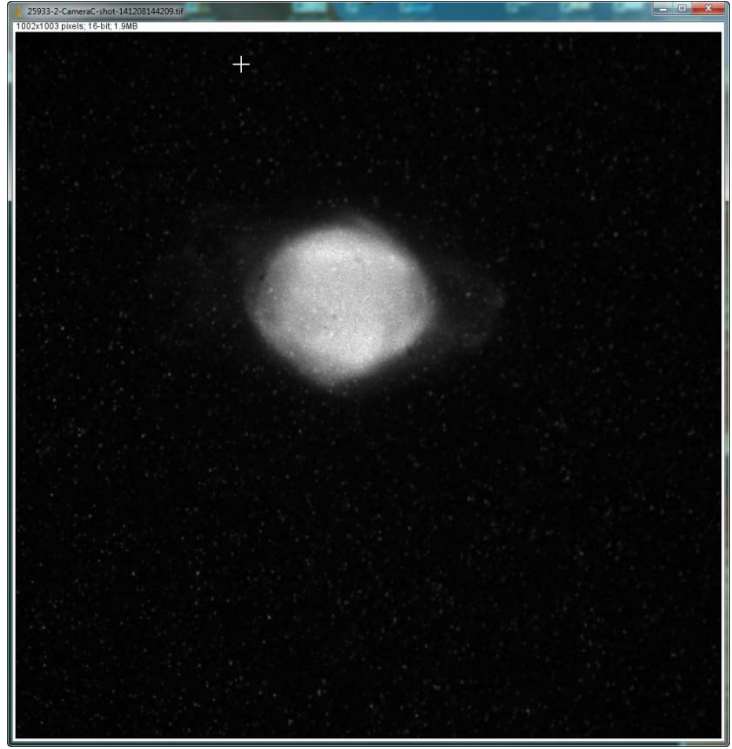

**Figure 25: Image of a kicked DARHT-II beam pulse in OTR light from a titanium foil. The MCP shutter gate for this image was 10 ns.**

### *Streak Cameras*

The biggest drawback to framing cameras is that they only capture the image at a single instant in time, or integrate the data over a window of time. To achieve time resolution one can use streak cameras. The trade-off is spatial resolution for temporal resolution. Since in practice spatial resolution is usually sacrificed by analyzing the image to obtain a single number quantifying its "size," the loss of spatial resolution is a small price to pay for the advantage of a continuous record in time.

The classical streak system collects light through a thin slit, the image of which is swept across the imaging plane. The resulting image is a slice of the object as it evolves in time. However, the DARHT-II streaks significantly improve on this by replacing the slit with an anamorphic lens system [57]. This compresses the light from the entire field of view into a line that is imaged onto a coherent, linear fiber-optic array that is cemented to the face of a remotely located streak camera.

Figure 25 shows a slit-less anamorphic streak of two anamorphic arrays arranged to orthogonally bisect a nominally round image of the beam from a Cerenkov target. Also shown is the projection of the streak images, also known as the line spread function (LSF). A significant advantage of this system is that it requires no alignment of physical slits to coincide with the expected position of the beam on the target. Moreover, it completely eliminates misinterpretation of an image caused by beam motion perpendicular to slit. Finally, anamorphic compression of the entire field of view in one direction simplifies the calculation of moments of the beam distribution, because the compression amounts to an optical integration in the direction orthogonal to the line image, which is in effect the line spread function (LSF) of the OTR source [58]. All that remains to be done is to compute the required moment of the LSF at each time. This results in an unambiguous, continuous record of the desired moment.

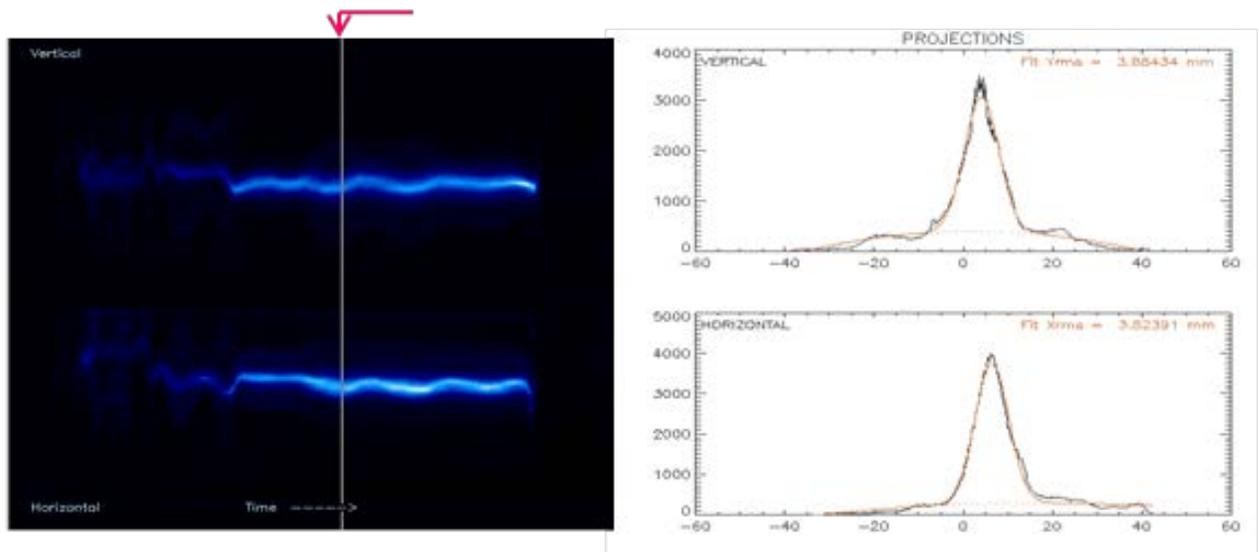

**Figure 26: (Left) Two-view streak image, showing vertical and horizontal views, with time running left to right. (Right) Projections of the streak images at the time indicated by the red arrow, with fit to data shown in red. These data clearly delineate the presence of a large halo for this particular tune.**

The beam imaging parameter of interest in the analysis of most experiments is its size, and the streak camera display is the most effective means for obtaining a continuous record of this during a single pulse. For example,

Figure 26 shows the temporal evolution of the rms width of the projections of the streak images in Figure 25. Since $y_{rms}$ is the beam "size" used to deduce the beam emittance, a temporal record of $y_{rms}$ during the pulse provides quantitative information about the uncertainties of this technique due to beam size variability.

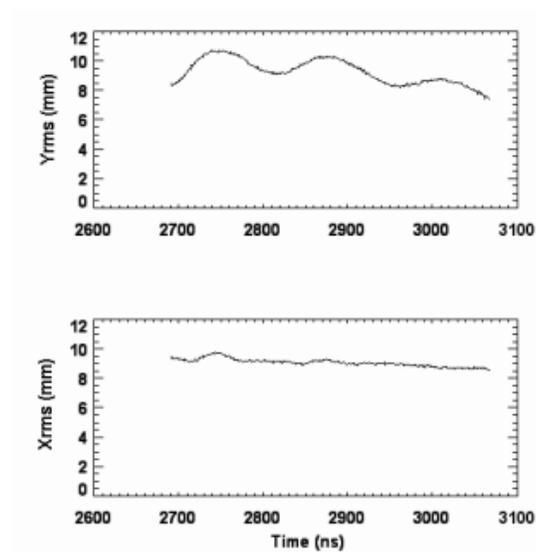

Figure 27: Rms width of projections (LSFs) of streak images in Figure 25.

More detailed information about the beam distribution can be obtained from higher order moments of the images, if desired. For example, Figure 27 shows the kurtosis of the streak images, which is a quantitative measure of the "peakiness" of the distribution based on fourth order moments. These moment analyses clearly show periodic variations of the beam halo evident in Figure 25. This effect could be easily missed if one were to rely solely on framing cameras.

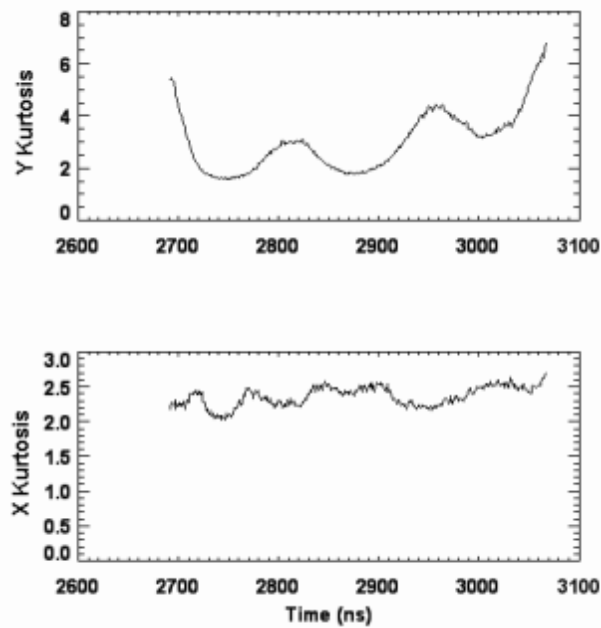

**Figure 28: Kurtosis of the projections of streak images in Figure 25.**

On DARHT-II, two slit-less anamorphic systems provide simultaneous projections in the horizontal (X) and vertical (Y) directions, which are recorded on a 1024x1024 CCD readout camera. Another orthogonal pair is oriented at 45 degrees in order to unambiguously resolve ellipticity. The images from this 4-view system can be tomographically reconstructed to provide more information about the beam shape. An example is shown in Figure 28 where reconstructed streak data is compared with PiMax framing images taken at the same time in light from the reverse side of the imaging target [59, 60].

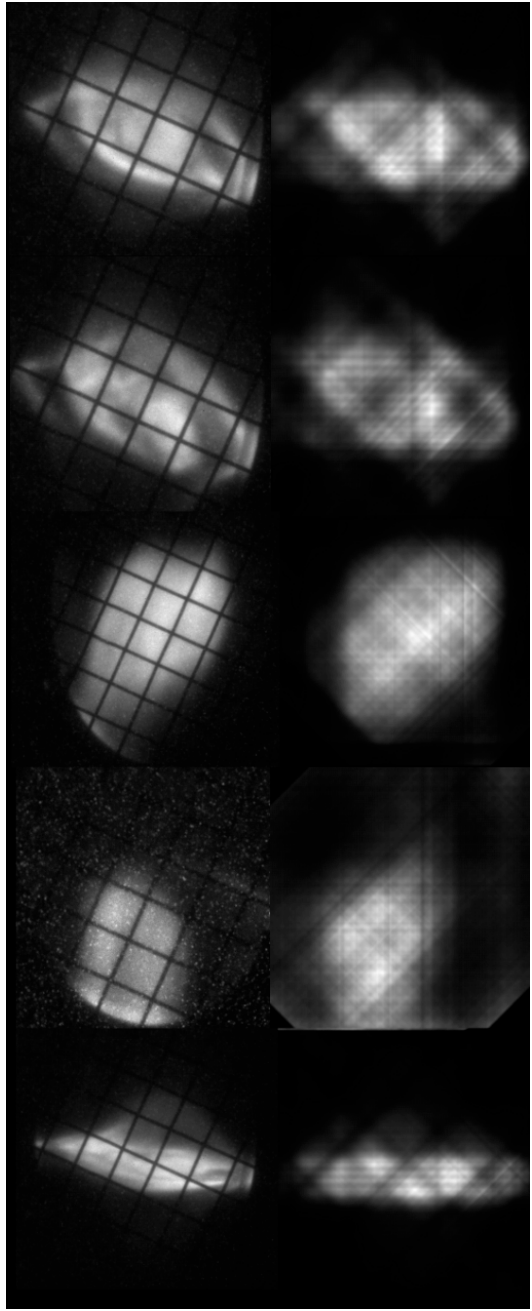

**Figure 29: Left column: PiMax images of beam that was deliberately mistuned through the DARHT-II quadrupoles in order to accentuate azimuthal asymmetry. Right column: Tomographic reconstruction of 4-view streak images at the same time a PiMax frames. PiMax and 4-view light came from opposite sides of the imaging target, so the screen grids are different in the two views.**

To summarize, the anamorphic optical system simplifies alignment, eliminates ambiguity resulting from beam motion, and eases analysis to find continuous temporal records of beam parameters of interest. Tomographic reconstruction of images provides even more information about the shape of the beam, and can be made into a movie showing how the shape evolves with time.

## *Magnetic Spectrometers*

Two electron spectrometer concepts are being designed for the Scorpius project. Each spectrometer will measure electron kinetic energies centered on 2 MeV and 20 MeV, which allow for measurements at the injector and at end of the accelerator, respectively.

The Enhanced Capability for Subcritical Experiments (ECSE) project will provide flash radiography capabilities using an electron Linear Induction Accelerator (LIA) that is similar to the accelerators at the Dual Axis Radiographic Hydrodynamic Test (DARHT) facility. The strict requirements for flash radiography require a detailed understanding of the LIA's performance, including precision measurements of the electron beam energy. Therefore, two electron spectrometers are being built to measure electron kinetic energies at the injector and after the LIA with energies centered on 2 MeV and 20 MeV respectively. Each spectrometer will follow a design similar to the latest DARHT electron spectrometers, which have electron energies centered on 3.6 MeV and 16.4 MeV.

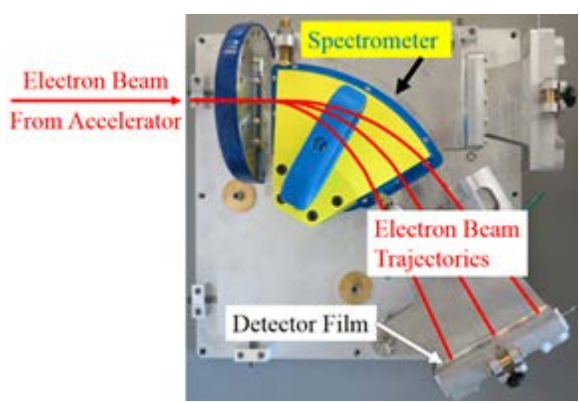

**Figure 30: Magnetic spectrometer setup showing trajectories of electrons with three different energies.**

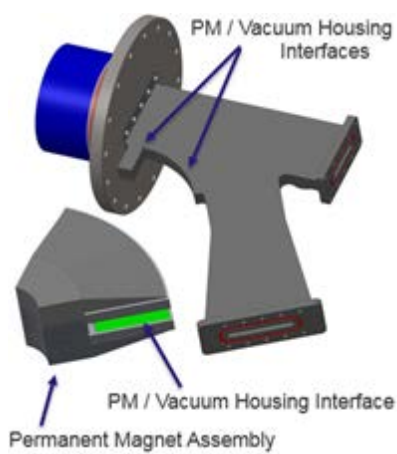

**Figure 31: Drawing of spectrometer showing permanent magnet separated from the vacuum chamber.**

The set-up is shown in Figure 29. A custom drift tube allowing for a 60° deflection has been designed to support both electron spectrometers and will replace sections of the LIA beam line during measurements. The spectrometers will slide on and off the drift tube as shown in Figure 30, and a system of guide bumpers (kinematic locating stops) will ensure reproducible mounting.

A narrow beam is required to provide accurate measurements, so two apertures will be mounted together to the front of the drift tube. The first aperture is a 76 mm thick graphite cylinder with a 2 mm diameter hole. The second aperture is a 25 mm thick tungsten cylinder with a 1 mm diameter hole. The emerging electron beam will therefore be 1 mm in diameter, and will pass through the magnetic field of the spectrometer before impinging onto a 100 mm wide detector film.

The required peak magnetic field strengths to deflect the 2 MeV and 20 MeV electrons are 0.044 T (440 G) and 0.367 T (3670 G), respectively. The 0.367 T field will be produced using a total of four permanent magnets to cover the top and bottom of the spectrometer's gap. The 0.044 T field will be produced by numerous smaller magnets which will be spaced apart from one another, and iron plates will cover the top and bottom of the spectrometer's gap to flatten the field. Examples of the magnetic field produced for the DARHT electron spectrometers with energies centered on 4 and 16 MeV are shown in Figure 2.

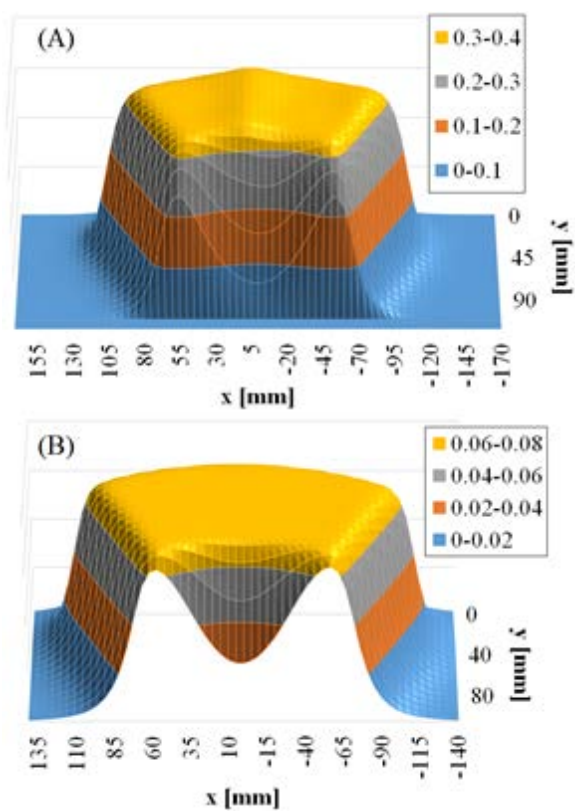

**Figure 32: Measured magnetic fields of the DARHT permanent magnet spectrometers. (A) 16-MeV magnet. (B) 4-MeV magnet.**

The magnetic field map for the 4 MeV DARHT spectrometer was scaled and the particle tracking computer code, General Particle Tracer (GPT), was used to determine the performance of the 2 MeV and 20 MeV electron spectrometers. GPT predicts that the 2 MeV spectrometer will be able to measure electrons with kinetic energies ranging from 1.7 MeV to 2.4 MeV which span ~10 cm in the imaging plane, while the 20 MeV spectrometer will be able to measure electrons with kinetic energies ranging from 17.2 to 23.7 MeV, which also span ~ 10 cm in the imaging plane. Therefore, 1-mm resolution by the imaging camera should result in energy resolution better than 0.5%.

Calibration of the spectrometers will be performed using precision ion beams such as those used to calibrate the spectrometer that has provided all DARHT beam energy data to date.

## B. Non-Invasive Diagnostics

### Beam Position Monitors

Beam monitors for measuring beam position and current will be based on proven DARHT designs. Each beam position monitor (BPM) will be an array of four $B_\theta$-field detectors spaced 90°apart. Each detector will be a balanced, shielded-loop design with a Moebius crossover to minimize common mode signals arising from ground loops, radiation driven Compton currents, direct beam spill pickup, electric field pickup, and/or other interfering noise, EMP, or backgrounds. As on DARHT-II, each shielded-loop will be formed from semi-rigid sub-miniature coax. The Moebius crossover is made by soldering the center conductor of one half of the loop to the outer conductor of the other half. The output signals from each half are proportional to almost exactly the same sensing area, and of opposite polarity. Subtraction of (calibrated) signals from each side cancels common mode and gives the desired signals proportional to $dB_\theta/dt$, which are then integrated and further processed in software to give beam parameters. The integrated signals from the four detectors are summed to yield the beam current, and opposing pairs are differenced to measure the beam centroid position.

Suppose that $[r, \theta]$ is the position of a filamentary element of the beam, and $[R_W, \theta_k]$ is the position of the $k^{th}$ detector of an array of $N$ magnetic field detectors equally spaced around the beam tube, which has radius $R_W$. For DARHT-II, $N=4$. These detectors are oriented to be sensitive to $B_\theta$, and their signals are digitally recorded, then analyzed with software. (Alternative hardware implementations suitable for direct analog recording are discussed in Attachment A)

The analysis of the signals obtained with this array begins with integration of the raw data (which is proportional to $dB_\theta / dt$), and multiplication by calibration factors as required producing a data record for each detector equal to the azimuthal magnetic field, $B_\theta(R_W, \theta_k)$. To resolve the $m^{th}$ azimuthal harmonic of the field these records are multiplied by either $\sin m\theta_k$ or $\cos m\theta_k$ and summed to yield:

$$\sigma_m^s = \sum_{k=1}^{N} B_\theta\left(R_W, \theta_k\right) \sin m\theta_k \quad , \tag{20}$$

and

$$\sigma_m^c = \sum_{k=1}^{N} B_\theta\left(R_W, \theta_k\right) \cos m\theta_k \quad . \tag{21}$$

As shown in Ref. 1, these sums are related to the position of a single current element of the distribution located at $[r, \theta]$ by

$$\sigma_m^s = N \frac{\mu_0 i}{2\pi R_W}\left(1 + \varepsilon_{m,N}^s\right)\rho^m \sin m\theta \quad , \tag{22}$$

and

$$\sigma_m^c = N \frac{\mu_0 i}{2\pi R_W}\left(1 + \varepsilon_{m,N}^c\right)\rho^m \cos m\theta \quad . \tag{23}$$

Here $i$ is the filamentary current, $\rho^m = \left(r / R_W\right)^m$, and the $\varepsilon_{m,N}^{s,c}$ are small aliasing errors resulting from the discrete nature of the detector array. The aliasing errors become smaller as the number of detectors increases. By ignoring, for the moment, the aliasing errors, Eq. (22) and (23) can be used as Green's functions to find the moments of the current distribution. The first ($m=1$) harmonic is required for determination of beam position.

Now, consider the current-distribution-weighted mean of any function;

$$\langle f(x,y)\rangle = \frac{\iint f(x,y)j(x,y)\,dx\,dy}{\iint j(x,y)\,dx\,dy} = \frac{1}{I}\iint f(x,y)j(x,y)\,dx\,dy \,. \quad (24)$$

Next, Equations (22) and (23) can be integrated over the current cross-section with $m = 1$.

$$\Sigma_1^s = N\,\frac{\mu_0 I}{2\pi R_W}\,\frac{\langle x\rangle}{R_W}\quad, \qquad\qquad (25)$$

$$\Sigma_1^c = N\,\frac{\mu_0 I}{2\pi R_W}\,\frac{\langle y\rangle}{R_W}\quad, \qquad\qquad (26)$$

Here we have used the notation

$$\Sigma_m^{c,s} = \left\langle \sigma_m^{c,s}\right\rangle\quad.$$

Equations (25) and (26) are supplemented by the direct, unweighted sum of all $N$ detectors;

$$\Sigma_0 = N\,\frac{\mu_0 I}{2\pi R_W}\quad, \qquad\qquad (27)$$

which is used as a normalization factor to finally obtain:

$$\langle x\rangle = R_W\,\frac{\Sigma_1^s}{\Sigma_0}\quad, \qquad\qquad (28)$$

$$\langle y\rangle = R_W\,\frac{\Sigma_1^c}{\Sigma_0}\quad, \qquad\qquad (29)$$

These two equations give the position of the center of the beam current.

The B-dot detectors for DARHT-II are of a balanced design to provide protection from common mode signals, such as radiation driven Compton currents, direct electron impact, electrostatic pickup, and ground loops. The latter has never been a problem for previous LIA accelerators because the pulselengths were so short that the accelerators were effectively transit time isolated from ground. However, ground loops are a problem for DARHT-II with its 2 microsecond-long pulse.

The balanced design b-dot loops have two output voltage signals, one equal to the rate of change of flux plus any common mode excited at the loop, the other equal to the negative of the change of flux plus any common mode. Thus, subtracting the two signals results in twice the rate of change of flux, and the common mode cancels out.

A "Moebius crossover" design is used, primarily because the background resulting from direct electron impact is lower for this type of balanced loop than for others. In this design the loop is formed from two lengths of semi-rigid coax, with the center conductor of each soldered to the outer conductor of the other midway around the loop, as shown in Figure 32. This type of balanced loop also has a safety advantage in that there is no danger of charged signal cables, because the center conductor is DC shorted to ground.

The loop emf generated by a time varying magnetic field linking the sensing area formed by the loop and back wall is just equal to the time derivative of the magnetic flux normal to the sensing area, $d\Phi/dt$. So, the output voltage signal from loop A is $A = d\Phi/dt + V_{CM}$, where $V_{CM}$ is any common mode, while the signal from the opposite sense loop B is $B = -d\Phi / dt + V_{CM}$. Subtracting the two signals eliminates the common mode.

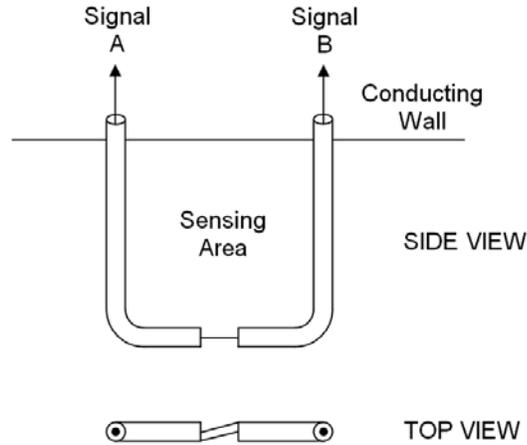

**Figure 33: Balanced "Moebius crossover" B-dot detector, showing how center conductors are soldered to outer conductors of opposing sides to form full flux sensing loops.**

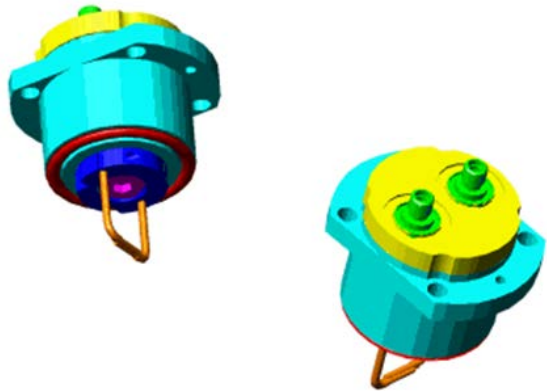

**Figure 34: Balanced B-dot detectors constructed for DARHT-II.**

The DARHT-II detectors incorporate all metal vacuum seals, and welded vacuum signal feedthroughs (SMA). There are no organics exposed to the vacuum. The loops are constructed from glass insulated semi-rigid coaxial cable that is formed under heat. The design of the individual detectors, and the BPM ring in which they are housed, minimizes time varying effects resulting from magnetic field soak during the long, 2-μs DARHT pulse.

Finally, signals from adjacent pairs of detectors can be differenced to monitor second-order azimuthal asymmetry (beam ellipticity). The quadrupole moment $Q$ of the beam current distribution, $Q^2 \equiv \left| \det Q^{ij} \right|$, is a measure of the ellipticity of the distribution. Here, the components of the quadrupole tensor of the current distribution are given by $Q^{ij} = \iint j(x, y)(2x_i x_j - \delta_{ij} r^2) dx dy / I$, where $j$ is the current density and $I$ is the total current. However, unambiguous measurement of $Q$ for a tilted elliptical distribution requires an eight-detector BPM. Four-detector BPMs can only be used to make an initial survey of where the beam is out of round ($Q \neq 0$).

Calibration of the current and position sensitivity of each BPM will be accomplished in a "coaxial" test stand with an inner conductor that can be accurately offset from center. These calibrations enable us to make measurements of beam current with less than +/- 1% uncertainty, and position to better than +/- 0.5%, which include the inaccuracies of all BPM system components including data recording.

## RF detectors

Each BPM will include a fifth $B_0$-field detector designed to provide independent high-bandwidth monitoring of RF beam motion produced by BBU. With this diagnostic, the growth of BBU through the LIA can be monitored on every shot, and the tune adjusted to reduce it to acceptable levels without lengthy data analysis.

## Diamagnetic Loops

In Scorpius the beam will be created by a cathode which has no magnetic flux linking it, so it has no magnetic field angular momentum. Since it has no azimuthal motion at the cathode surface, its total (canonical) angular momentum is also zero. When such a beam enters a magnetic field it acquires a rotation interaction with the radial fringe field. This motion generates an azimuthal current which produces an axial magnetic field inside the beam that opposes the external field. This is the source of beam diamagnetism. For short pulses, such as in Scorpius and DARHT, total axial magnetic flux is conserved inside of the conducting beam pipe, so the diamagnetic decrease of field inside the beam is accompanied by an increase in field outside of the beam (see Figure 34 ). The excluded flux can also be detected by a loop that surrounds the beam, but is located within the conducting outer wall [61, 62, 63, 64]. This is known as a diamagnetic loop (DL).

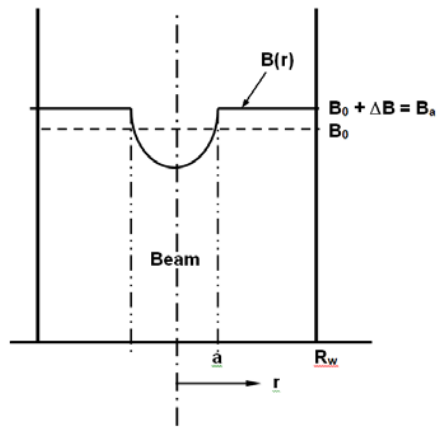

**Figure 35: Axial magnetic field inside of flux conserving beam pipe. $B_0$ refers to initial field produced by focusing solenoids, which is increased by flux excluded from beam by its diamagnetism.**

For a total change of flux $\Delta\Phi$ measured by the DL, it can be shown that the beam rms radius is given by

$$\pi R_{rms}^2 \approx -\frac{I_A}{I_b}\frac{\Delta\Phi}{B_0}\frac{1}{1-R_L^2/R_W^2} \tag{30}$$

where $I_A = 17.08\beta\gamma$ kA , $B_0$ is the initial solenoidal field, $R_L$ is the loop radius, and $R_W$ is the radius of the conducting wall. This equation shows that the sensitivity of the measurement ( $d\Delta\Phi/dR_{rms}$ ) is proportional to the area between the loop and the wall and also to the bias magnetic field $B_0$ . We have taken advantage of this in the design of a self-contained DL diagnostic incorporating its own solenoid so that the loop is centered on the region of maximum $B_0$ (maximum sensitivity), and which features a large area between the loop and the outer wall, as shown in Figure 35. Calibration of the loop is accomplished by inserting fast pulsed single-layer coils of various winding pitch and diameter [63, 64].

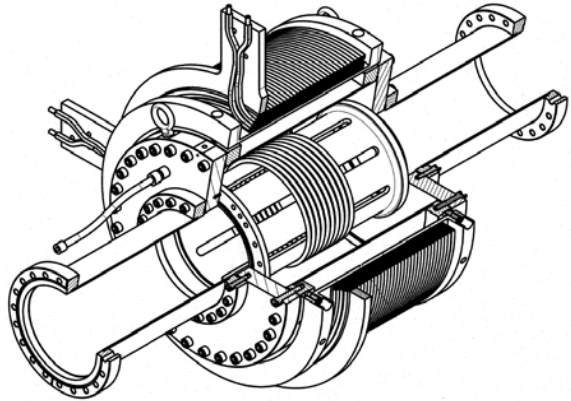

**Figure 36: Cutaway drawing of a prototype diamagnetic loop diagnostic that can be inserted into the Scorpius beamline. The multiturn DL is wrapped on the beam pipe, which is slotted to allow axial magnetic fields to penetrate. The solenoid is wrapped on the conducting outer cylinder, which provides the outer flux conserving surface defining the sensitivity to beam radius.**

Clearly, the ability to make beam size measurements on every shot without perturbing the beam would be a significant improvement over the present art, which require invasive beam imaging. Non-invasive beam size measurements should greatly improve the accuracy of emittance measurements using the focal scan technique, because it obviates the concern for imaging target damage which limits present measurements. Finally, it should be possible to incorporate these DL diagnostics throughout Scorpius, using the built-in solenoid as part of the tune, thereby providing simultaneous beam radius measurements throughout the accelerator on every shot. Figure 35

# VII.  Codes

The codes that were used to assess the Scorpius design are also in regular use for simulations of the DARHT accelerators, and wherever possible, they have been experimentally validated with data from those LIAs. Briefly, the codes used are

- XTR; a stationary beam envelope and centroid equation solver.
- LAMDA; a time resolved beam envelope and centroid equation solver.
- LSP-Slice; a particle-in-cell (PIC) code.

These codes require externally applied magnetic and electric fields as input. These fields were derived from electromagnetic simulations of the solenoid and accelerating gap of the Scorpius cell design (Figure 35).

## A. Electromagnetics

External electromagnetic fields required for the beam simulation codes were calculated using finite element models of the accelerator elements. The Field Precision Tricomp suite of codes [65, 66, 67] was used for this task.

Magnetic fields of the transport solenoids were calculated with the Field Precision code PerMag [65]. PerMag is a code for the design of electromagnets and permanent magnet devices. The program calculates magnetostatic fields in complex geometries with coils, linear or non-linear ferromagnetic materials, anisotropic materials and permanent magnets. The unitized, self-contained package addresses all aspects of the problem: mesh generation, finite-element solution, analysis and plotting. PerMag employs finite-element methods on variable resolution conformal triangular meshes for high accuracy and speed. The mesh size is limited only by the installed memory. The program handles three-dimensional cylindrical problems (symmetry in $\theta$) and two-dimensional rectangular problems (arbitrary variations in x and y with infinite extent in z).

Electric fields produced at the accelerating gaps were calculated by the Field Precision code Estat [65]. EStat is a code that calculates electrostatic fields in complex two-dimensional geometries. Simulated systems may include electrodes, conductors, dielectrics, and space-charge. The unitized, self-contained package addresses all aspects of the problem: mesh generation, finite-element solution, analysis and plotting. EStat employs finite-element methods on variable-resolution conformal triangular meshes for high accuracy and speed. Mesh and geometry limitations are the same as for PerMag. Analysis functions include a wide variety of $\phi$ and E plots as well as automatic calculation of Gaussian surface integrals, electrostatic energy and induced charge.

## *Magnetic Field Modeling*

The solenoidal magnetic fields used for the beam simulations are based on the Scorpius 4-pulse cell design, shown in Figure 36. The key enabling technology for a reliable multi-pulse LIA is the accelerating cell. For Scorpius, the cell design is based on the proven DARHT-I cell, but with the ferrite cores replaced with Metglas to provide enough flux swing (volt-seconds) for four pulse operation. With this design one need only reset the cores before each four-pulse burst.

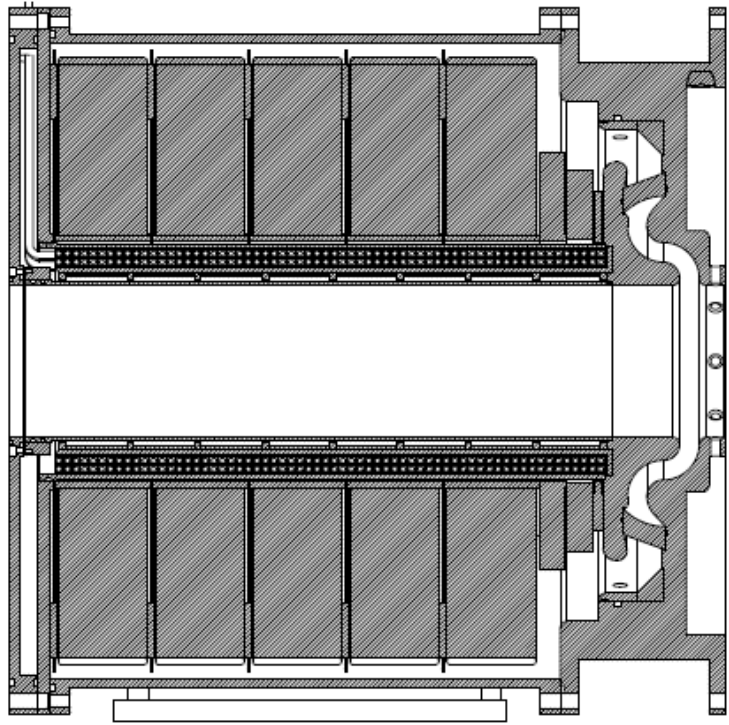

Fig. 1

**Figure 37: Scorpius cell design used for magnetic and electric field calculations. The cell design is based on the proven DARHT-I cell, but with the ferrite cores replaced with Metglas to provide enough flux swing (volt-seconds) for four pulse operation. With this design one need only reset the cores before each four-pulse burst.**

The solenoid is a double layer of square, hollow core conductor with 156 total turns. We used the Field Precision TriComp PerMag program [65] to calculate the field in the cell for a 100 A drive current. Figure 37 shows the resulting contours of $rA_\theta$, which is proportional to magnetic flux. The simulation indicates that this magnet will produce a peak field on axis of 3.18 Gauss/Amp. The field on axis simulated by PerMag was fit to the ideal current-sheet solenoid model used in our XTR and LAMDA envelope codes. The field on axis for such a solenoid is given by

$$B(z) = B(0) \left[ \frac{\left[ L^2/4 + R^2 \right]^{1/2}}{L} \right] \left\{ \frac{L/2 + z}{\left[ \left( L/2 + z \right)^2 + R^2 \right]^{1/2}} - \frac{L/2 - z}{\left[ \left( L/2 - z \right)^2 + R^2 \right]^{1/2}} \right\} \quad (31)$$

The best fit of the PerMag results to this model gives $L = 51.48$-cm effective length, $R = 8.955$-cm effective radius, and $B(0) = 3.18$ G peak field for a 1.0-A drive current. Figure 38 is a plot of the axial magnetic flux density, $B_z$, on axis as calculated by PerMag, along with the XTR model fit for comparison. Although the PerMag result is slightly left-right asymmetric due to the asymmetry of the magnetic material surrounding the solenoid, the symmetric ideal solenoid model used in XTR and LAMDA agrees well enough for the CDR simulations that follow.

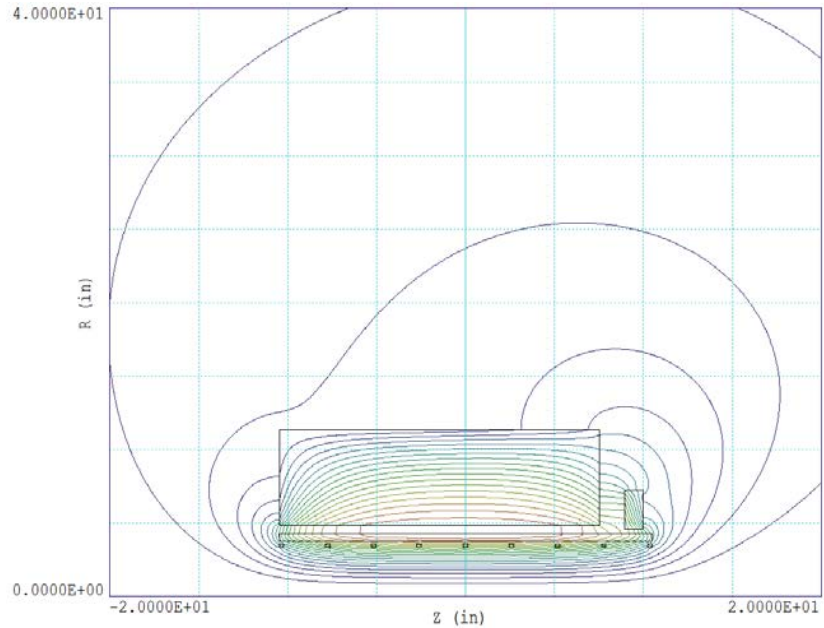

**Figure 38: Contours of $rA_\theta$ for the Scorpius test cell design shown in Fig 1.**

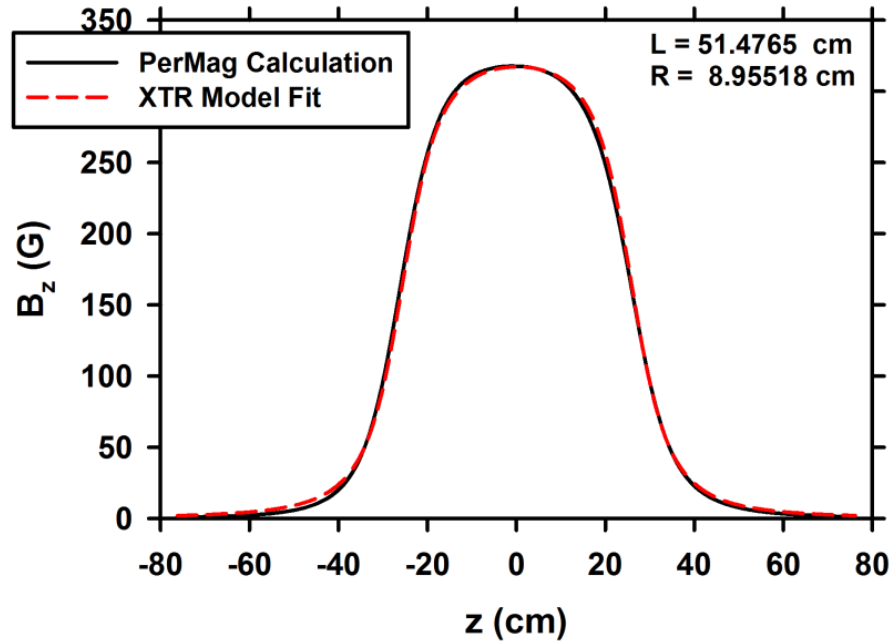

**Figure 39: Axial magnetic field on the axis of the Scorpius cell solenoid when powered with 100 A. (solid black line) Calculated with the TriComp PerMag code. (dashed red line) Fitted model used in XTR, LAMDA, And LSP simulations.**

## *Electric Field Modeling*

The XTR envelope code uses a thin Einzel-lens approximation for the acceleration and focusing of the beam by the LIA gaps, but the PIC code requires explicit $E_z$ on axis in tabular form. This table was generated from an electrostatic simulation of the gap region for the Axis-I geometry, which will be duplicated on Scorpius. The TriComp Estat code [66] was used for this calculation. Figure 39 shows the electrostatic potentials for this simulation with 250 kV across the gap. Only the features of the gap region that might affect the field on axis were included. Figure 40 shows the resulting field on axis, which was used with the gap locations to create an input file for LSP-Slice PIC simulations.

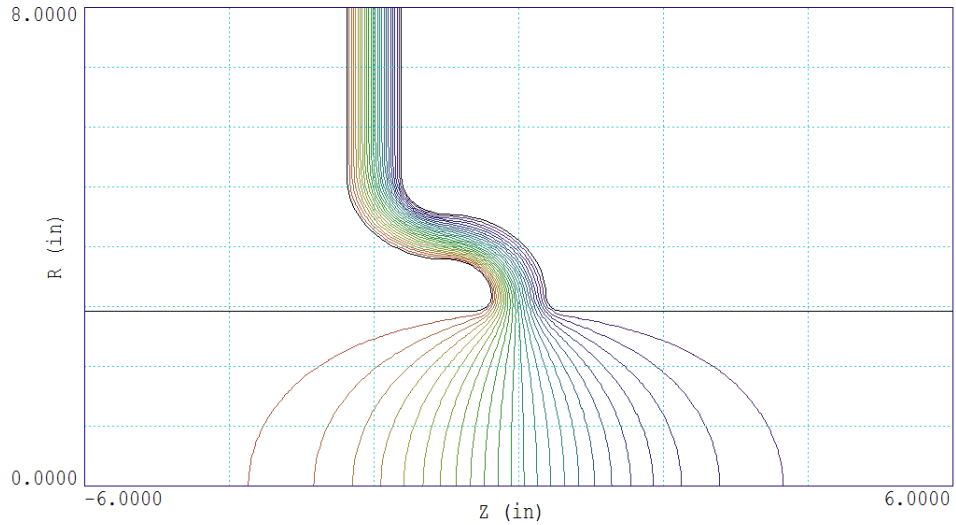

**Figure 40: Equipotentials of the accelerating electric field at 10-kV intervals in the region of the Scorpius gap for 250-kV gap voltage.**

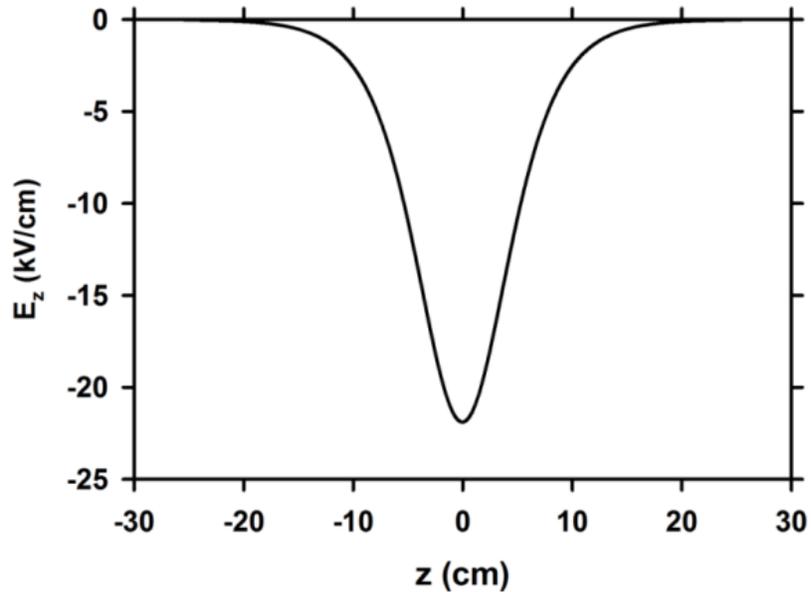

**Figure 41: Accelerating electric field on axis calculated by Estat for the Scorpius cell. This field is used in PIC code simulations of beam transport.**

## B. Envelope Codes

Design of tunes for the DARHT accelerators is accomplished using envelope codes. The two most frequently used are XTR and LAMDA. XTR was written by Paul Allison in the IDL language [4]. LAMDA was originally written by Tom Hughes and R. Clark [18]. In both of these codes the radius $r$ of a uniform density beam is calculated from an envelope equation [8]. In the DARHT accelerators the beam is born at the cathode with no kinetic angular momentum and a nearby reverse polarity solenoid to cancel out the magnetic flux at the surface. Thus, the beam has no canonical angular momentum, and the envelope equation is

$$\frac{d^2 r}{dz^2} = -\frac{1}{\beta^2 \gamma} \frac{d\gamma}{dz} \frac{dr}{dz} - \frac{1}{2\beta^2 \gamma} \frac{d^2 \gamma}{dz^2} r - k_\beta^2 r + \frac{K}{r} + \frac{\varepsilon^2}{r^3}$$ (32)

It can be shown that this same equation holds true for any axisymmetric distribution [68], so long as the radius of the equivalent uniform beam is related to the rms radius of the actual distribution by $r = \sqrt{2} R_{rms}$ . Here, $\beta = v_e / c$ , $\gamma = 1/\sqrt{1-\beta^2}$ , are the usual relativistic parameters, and the beam electron kinetic energy is $KE = (\gamma - 1) m_e c^2$ . The betatron wavelength is

$$k_\beta = \frac{2\pi B_z}{\mu_0 I_A}$$ (33)

where $I_A = 17.08 \beta\gamma$ kA , and the generalized perveance is $K = 2I_b / \beta^2 \gamma^2 I_A$ . The emittance which appears in Eq. (32) is related to the normalized emittance by $\varepsilon = \varepsilon_n / \beta\gamma$ , where

$$\varepsilon_n = 2\beta\gamma \sqrt{\langle r^2 \rangle \left[ \langle r'^2 \rangle + \langle (v_\theta / \beta c)^2 \rangle \right] - \langle rr' \rangle^2 - \langle rv_\theta / \beta c \rangle^2}$$ (34)

which is invariant through the accelerator under certain conditions.

We sometimes say that the beam is either space-charge dominated or emittance dominated, depending on the relative size of the last two terms on the right-hand side. These are both defocusing, and are equal when $r^2 = \varepsilon^2 / K = (8.5 \text{kA} / I_b) \varepsilon_n^2 \beta\gamma$ . This shows that high energy beams and small beams tend to be emittance dominated, which greatly simplifies solution of the envelope equation.

## *XTR*

The XTR envelope code [4] was used to develop the tune for the Scorpius accelerator conceptual design shown in Figure 3. The simple envelope equation in Eq. (32) is further improved in XTR as follows. The energy dependence of the beam due to the gaps is approximated by a linear increase in $\gamma$ accompanied by a thin-einzel-lens focus. Between gaps $\gamma$ used in Eq. (32) is the value at the beam edge, which is space-charge depressed by $\Delta\Phi \approx 30 I_b (2 \ln R_w / r)$ , where $R_w$ is the radius of the beam pipe [69]. XTR also uses the magnetic field at the beam edge, including a first order approximation to account for the flux excluded by a beam rigidly rotating in the magnetic field due to the invariance of canonical angular momentum.

## *LAMDA*

LAMDA (an acronym for Linear Accelerator Model for DARHT) is a transport code which advances the beam centroid and envelope in an induction accelerator from the injector to the final focus region. The code can treat an entire beam pulse, or just one beam-slice. LAMDA computes the effects of

- magnet misalignments,
- background gas ionization,
- gap voltage fluctuations,
- beam breakup and image-displacement instabilities,
- resistive-wall instability,

- ion hose instability.

In LAMDA, the beam is modeled as a string of n rigid disks (Figure 41 ). For a circular beam, the parameters associated with each disk are the normalized current, energy , transverse centroid displacements , and radius. Each of these five quantities is a function of z, the axial distance along the accelerator, and the time measured from the head of the pulse. External Lorentz forces due transverse magnetic fields, gaps, and walls are applied to each disk.

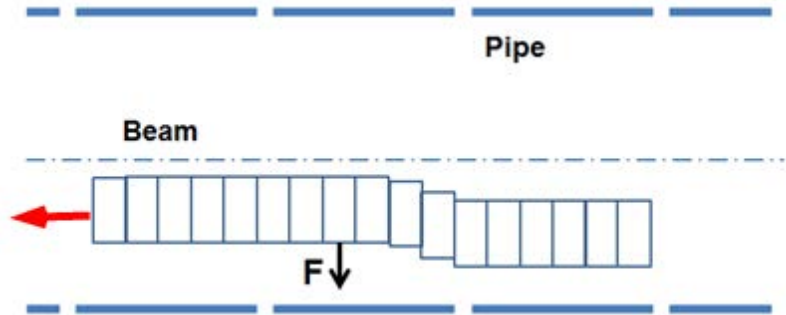

**Figure 42: Beam pulse is represented by n disk beam disks. External Lorentz forces due transverse magnetic fields, gaps, and walls are applied to each disk.**

It is assumed that there is no interaction between the beam disks, except at gaps (or due to wall resistivity, which is treated similarly). As a result, all the beam disks can be transported together; i.e., at any point in the simulation, all of the disks are at the same axial position. This simplifies book-keeping in the code. In general, there are two types of differential equations to be solved for the beam motion: the equations for the spatial motion of each beam disk along the accelerator, and the equations for the temporal variation of the voltage and BBU forces at the gaps.

### *Centroid Equations of Motion*

To model the trajectory of the beam centroid, LAMDA solves the Lorentz force equation for each disk

$$
\begin{aligned}
\frac{dp_x}{dt} &= e\left( E_x - \frac{v_z}{c}B_y + \frac{v_y}{c}B_z \right) \quad, \\
\frac{dp_y}{dt} &= e\left( E_y + \frac{v_z}{c}B_x - \frac{v_x}{c}B_z \right) \quad,
\end{aligned}
\tag{35}
$$

and transforms the independent variable from time ($t$) to position ($z$) in the lab frame

$$\gamma\beta^2 x'' = -\gamma' x' + \frac{e}{mc^2}(E_x - \beta B_y + \beta y' B_z) \quad ,$$

$$\gamma\beta^2 y'' = -\gamma' y' + \frac{e}{mc^2}(E_y + \beta B_x - \beta x' B_z) \quad ,$$

(36)

where the prime symbol denotes $d/dz$, $\beta = v_z/c$, and $\gamma = \sqrt{1-1/\beta^2}$. The electromagnetic fields in these equations include all external fields (solenoids and gaps) plus the fields of the beam image in the beam pipe and the fields resulting from the curvature of the centroid trajectory [32].

## BBU Algorithm

The approach taken by the LAMDA BBU algorithm is to calculate the kick given to a beam disc using a wake function describing the electro-magnetic fields generated by the preceding disks interacting with the cavity [51]. This can be expressed as a convolution integral

$$\frac{\triangle p_x}{mc} = \int_0^t w(t-t')\iota(t')\xi(t')dt' \quad ,$$

(37)

where $\iota$ is the current normalized to 17.05 kA, $\xi$ is the centroid displacement in cm, and the wake function $w$ has units of cm$^{-2}$. In LAMDA, the right-hand side of Eq. (37) is actually solved using the Fourier convolution theorem and fast Fourier transforms. That is

$$\int_0^t w(t-t')\iota(t')\xi(t')dt' = \int_{-\infty}^{\infty} Z(\omega)F(\omega)e^{-i\omega t}d\omega \quad ,$$

(38)

where $F(\omega)$ is the Fourier transform of $\iota(t)\xi(t)$, and $Z(\omega)$ is the Fourier transform of the wake function $w(t)$. $Z(\omega)$ is a complex quantity, commonly known as the transverse coupling impedance, and in LAMDA it is represented by a multiple resonance model due to Briggs [70, 71]

$$Z(\omega) = 120\left(\frac{d}{b^2}\right)\left[\sum_n\left(\frac{\eta_n}{1+i2Q_n(\omega/\omega_n-1)} - \frac{\eta_n}{1+i2Q_n(\omega/\omega_n+1)}\right) - i\eta_0\right] \quad ,$$

(39)

where $b$ is the beam pipe radius, $d$ is the accelerating gap width, $\eta_n$ is a cavity form factor for the resonance at frequency $\omega_n$, and $Q_n$ is the cavity quality factor for the $n$th resonance. If the units of the pipe radius ($b$) and accelerating gap ($d$) are in cm, the coupling impedance ($Z$) has units of Ohm/cm. In practice, this form is fit to experimental data, such as in [70], or to the results of an electromagnetic code, such as AMOS [72]. We fit the function to experimental data, using DARHT-I data, since we propose to replicate the DARHT-I cavity geometry on Scorpius.

## Resistive-wall Algorithm

For a constant-current coasting beam the time-varying electromagnetic fields produced by the conducting-wall images of a beam displaced a distance $\xi(t)$ from the centerline were derived in [40]. These fields are proportional to $\xi(t)$, and for early times, the resulting radial force on the beam toward the wall is approximately

$$\mathbf{F}(z,t) = \mathbf{e_x}\left[\frac{2eI}{b^2\beta c\gamma^2}\xi(z,t)\right.$$
$$\left. + \frac{4eI\beta}{\gamma\pi b^3\sqrt{\sigma}}\int_{-\infty}^{t}\frac{d\xi}{dt'}\sqrt{t-t'}dt'\right]$$

(40)

where $e$ is the electron charge, $I$ is the beam current, $b$ is the beam pipe radius, $\beta\gamma = \sqrt{\gamma^2-1}$ is the normalized electron momentum [41]. The convolution integral in Eq. (40) is calculated using a discrete convolution approximation, the discrete convolution theorem, and FFTs, exactly as for the BBU model.

*Corkscrew Motion*

LAMDA calculates the transverse fields due to cell misalignments from first-order expansions of the off-axis fields of the solenoids. For example, the transverse fields for 0.29-mm rms offsets for the Scorpius CDR tune are shown in Figure 11. These fields are used with the centroid equations of motion to calculate the trajectory of each beam disk, which can have different energies depending on the temporal variation of the accelerating potentials. Thus, each beam disk can have a different trajectory, constrained by the solenoidal focusing field, which results in the characteristic corkscrew motion.

# C. Particle-in-Cell (PIC) codes

## *LSP*

The LSP-slice algorithm is based on the LSP PIC code [73]. A slice of beam particles located at an incident plane of constant z are initialized on a 2D transverse Cartesian ($x, y$) grid. The use of a Cartesian grid admits non-axisymmetric solutions, including beams that are off axis. Simulations were performed on a workstation with 32 processors. Multiprocessing reduced the time for a typical run from the more than 30 hours required for earlier single processor runs to less than 4 hours. For axisymmetric beams, one can use a faster version of the code based on a 1D cylindrical grid. Using all 32 processors, the typical 1D run completes in less than 4 minutes. Excellent agreement between the 2D and 1D results have been obtained in comparison tests.

Initial electro- and magneto-static solutions are performed prior to the first particle push to establish the self-fields of the beam, including the diamagnetic field if the beam is rotating. After this initialization step, Maxwell's equations are solved on the transverse grid with $\partial/\partial z = 0$, and then the particles are pushed by the full Lorentz equations. At each time-step the grid is assumed to be located at the axial center-of-mass of the slice particles $z(t)$, which is propagating in the $z$ direction.

The initial particle distribution of the slice is extracted from a full $x, y, z$ LSP simulation. The distribution is a uniform rigid rotor with additional random transverse velocity. The rotation is consistent with zero canonical angular momentum in the given solenoidal magnetic field at the launch position. The random transverse velocity is consistent with the specified emittance. Best agreement between LSP-slice and full LSP 2D simulations was obtained when the slice model is initiated at an envelope extreme, where the beam convergence is zero, so this condition was used for all simulations for this article. Also, for this article, 2D simulations used 70,688 particles, and 1D simulations used 4,000 particles.

External fields are input as functions of $z$, and are applied at the instantaneous axial center-of-mass location. External fields that are azimuthally symmetric (fields from solenoids and gaps) are input as on-axis values, and the off-axis components are calculated up to sixth order using a power series expansion based on the Maxwell equations [74]. In this way the nonlinearities of the accelerator optics are included in the slice simulations. The on axis magnetic field input was obtained from the XTR simulation shown in Fig. 1. Transverse magnetic fields from steering dipoles and cell misalignments were input as $x, y$ values that uniformly fill the solution space, an approximation that is obviously best for a beam near the axis. These dipole fields were obtained from XTR, which calculates them on axis from steering dipole excitation currents and cell misalignments, which have been measured [75, 76].

Although the envelope equation only deals with axisymmetric beams centered on axis, the concept of beam emittance is much more general, and it can be calculated for non-axisymmetric distributions in LSP-slice simulations. Consider a non-rotating beam with normalized distribution $\rho(x, x')$ in the $(x, x')$ plane of phase space. The position of the centroid of this distribution is at

$$
\begin{aligned}
\langle x \rangle &= \iint x \rho(x, x) dx dx' \\
\langle x' \rangle &= \iint x' \rho(x, x) dx dx'
\end{aligned}
\qquad .
\tag{41}
$$

Now consider the $2 \times 2$ matrix with elements defined by

$$
\begin{aligned}
\sigma_{xx} &= 4 \iint (x - \langle x \rangle)^2 \rho(x, x') dx dx' = 4 \langle x^2 \rangle - 4 \langle x \rangle^2 \\
\sigma_{x'x'} &= 4 \iint (x' - \langle x' \rangle)^2 \rho(x, x') dx dx' = 4 \langle x'^2 \rangle - 4 \langle x' \rangle^2 \\
\sigma_{x'x} &= 4 \iint (x - \langle x \rangle)(x' - \langle x' \rangle) \rho(x, x') dx dx' \\
&= 4 \langle xx' \rangle - 4 \langle x \rangle \langle x' \rangle
\end{aligned}
\qquad ,
\tag{42}
$$

with $\sigma_{xx'} = \sigma_{x'x}$. The sigma matrix for the beam is

$$
\sigma_x = \begin{pmatrix} \sigma_{xx} & \sigma_{xx'} \\ \sigma_{x'x} & \sigma_{x'x'} \end{pmatrix}
\qquad .
\tag{43}
$$

This matrix is related to the area occupied by the beam in the $x, x'$ cut through phase space by $A_x = \pi \sqrt{\det \sigma_x}$ [77]. Since the emittance is defined as $\varepsilon_{rms} = A / \pi$ in beam optics, it follows that the rms emittance in the $x, x'$ cut through phase space is $\varepsilon_{x, rms} = \sqrt{\det \sigma_x}$. Without loss of generality, one can center the beam in $x, x'$ space, and then from Eq. (43) one gets

$$
\varepsilon_{x, rms} = 4 \sqrt{\langle x^2 \rangle \langle x'^2 \rangle - \langle xx' \rangle^2}
\qquad ,
\tag{44}
$$

which is again the Lapostolle "4-rms" emittance [78]. Multiplying by $\beta \gamma$ gets the normalized emittance. The LSP emittance algorithm follows a suggestion by [79], and generalizes this approach to $\varepsilon_{rms} = (\det \sigma)^{1/4}$ [53], where $\sigma$ is the $4 \times 4$ matrix formed from 4D moments as in Eq. (41) and Eq. (42) permuted through all transverse coordinates $x, x', y, y'$. This convention for $\varepsilon_{rms}$ reduces to Eq. (44) for axisymmetric beams.

### *AMBER*

AMBER is a ParticleInCell (PIC) code which models the evolution of a representative slice of a relativistic electron beam in a linear accelerator. The beam is modeled as a steady flow and therefore no electromagnetic waves: all the fields (external and self-fields) are electrostatic and magnetostatic fields. The possible elements describing the accelerator lattice are solenoids, accelerating gaps, pipes and apertures. Several kinds of beam distribution can be loaded: KV, Gaussian, semi Gaussian, etc. Alternatively, the user can reconstruct (or load) a distribution from the output of another code, for example, an interface generating the beam distribution from output produced from EGUN or LSP codes is available as an option.